\documentclass[11pt,a4paper]{article}
\usepackage{jheppub}
\usepackage{etoolbox}
\usepackage{comment}
\usepackage{amsmath}
\usepackage{slashed}
\usepackage{youngtab}  %This package is used to efficiently generate Young Tableaux%
\usepackage{float}
\usepackage[export]{adjustbox}
\usepackage{amsfonts}
\usepackage{subfig}
\def\one{{\,\hbox{1\kern-.8mm l}}}
\newcommand{\Dslash}{\not{\hbox{\kern-4pt $D$}}}
\newcommand{\pdslash}{\not{\hbox{\kern-2pt $\partial$}}}

\newcommand{\eps}{\epsilon}

\newcommand{\La}{\Lambda}
\newcommand{\R}{\rho}

\newcommand{\Comment}[1]{{}}

\def\IZ{{\mathbb Z}}

\def\IC{{\mathbb C}}
\def\IO{{\mathbb O}}
\def\IP{{\mathbb P}}

\def\IR{{\mathbb R}}
\newcommand{\bc}{\begin{center}}
	\newcommand{\ec}{\end{center}}
\newcommand{\ba}{\begin{array}}
	\newcommand{\ea}{\end{array}}
\newcommand{\beq}{\begin{equation}}
	\newcommand{\eeq}{\end{equation}}
\newcommand{\bea}{\begin{eqnarray}}
	\newcommand{\eea}{\end{eqnarray}}
\newcommand{\bmx}{\begin{pmatrix}}
	\newcommand{\emx}{\end{pmatrix}}

\newcommand{\be}{\begin{equation}}
	\newcommand{\ee}{\end{equation}}

\newcommand{\la}{\lambda}

\newcommand{\s}{\sigma}

\newcommand{\del}{\partial}
\newcommand{\half}{{\frac{1}{2}\,}}
\newcommand{\tr}{{\rm tr}}

\newcommand{\eref}[1]{Eq.\,(\ref{#1})}

\def\IB{\relax{\rm I\kern-.18em B}}
\def\ID{\relax{\rm I\kern-.18em D}}
\def\IE{\relax{\rm I\kern-.18em E}}
\def\IF{\relax{\rm I\kern-.18em F}}
\def\II{\relax{\rm I\kern-.18em I}}
\def\IZ{\relax{\sf Z\kern-.35em Z}}
\def\Id{\relax{1\kern-.32em 1}}
\def\IG{\relax\hbox{$\inbar\kern-.3em{\rm G}$}}
\def\IR{\relax{\rm I\kern-.18em R}}

\def\ket#1{| #1 \rangle}
\def\bra#1{\langle #1 |}

\def\vev#1{\langle #1 \rangle}
\title{Target space entanglement in Matrix Models}
\author[a]{Harsha R. Hampapura}
\author[a]{Jonathan Harper}
\author[a]{Albion Lawrence}

\affiliation[a]{Matrin A.Fisher School of Physics,\\
	Brandeis University, Waltham MA 02453, USA}

\preprint{BRX-TH-6658}
\emailAdd{hrharsha@brandeis.edu}
\emailAdd{albion@brandeis.edu}
\emailAdd{jharper@brandeis.edu}

\abstract{We study target space entanglement in gauged multi-matrix models as models of entanglement between groups of D-branes separated by a planar entangling surface, paying close attention to the implementation of gauge invariance. We open with a review of target space entanglement between identical particles, which shares some important features (specifically a gauged permutation symmetry) with our main problem. For our matrix models, we implement a gauge fixing well-adapted to the entangling surface.  In this gauge, we map the matrix model problem to that of entanglement of a $U(1)$ gauge theory on a complete or all-to-all lattice. Matrix elements corresponding to open strings stretching across the entangling surface in the target space lead to interesting contributions to the entanglement entropy.}
\begin{document}
\maketitle	
\flushbottom
\section{Introduction}  \label{sec: Introduction}
The study of quantum entanglement between spatial regions in strongly coupled, large-N theories, starting with the classic work  \cite{Ryu:2006bv,Ryu:2006ef,Hubeny:2007xt}, has led to crucial insights regarding the emergence of the bulk spacetime in 
holographic theories. However, we expect that other means of separating degrees of freedom and computing their entanglement 
are crucial for a variety of reasons:
\begin{itemize}
\item The existence of {\it entanglement shadows} (see for example \cite{Balasubramanian:2014sra,Freivogel:2014lja,Andy:2017}), regions of
the bulk that are not in the entanglement wedge of any spatial region.
\item The fact that the entangling surfaces for spatial entanglement in holographic conformal field theories do not resolve the large ``internal" manifolds transverse to the anti-de Sitter factors of the gravitational duals. Possible duals of extremal surfaces in this direction have
been discussed in \cite{MollabashiNoburoTadashi:2014,KarchUhlemann}. Note that these surfaces are not minimal \cite{Andy:2017}). 
\item The existence of large-$N$ gauged matrix quantum mechanics models such as \cite{BFSS:1997,BMN}\ with holographic duals, and no spatial
regions to entangle. For the one-matrix model a direct calculation with a clear dual interpretation was provided in \cite{Das:1995vj,Hartnoll:2015}, and there has been some discussion of bulk surfaces and possible dual boundary entanglement quantities in the BFSS model \cite{Anous_2019,Das:2020jhy,Das:2020xoa}, but the question remains open.\footnote{See the end of this section for a discussion of the relation between this paper and the latter two references.}
\end{itemize}

A natural avenue to investigate is the entanglement between matrix degrees of freedom in large-N gauged matrix quantum theories. It has been argued for many years that these degrees of freedom are crucial in exploring local physics at scales below the bulk radius of curvature. An early hint comes from \cite{SusskindWitten:98}, from which it is clear that in global anti-de Sitter space, a patch of size $R_{AdS}$ has $N^2$ degrees of freedom, when the dual holographic theory is a large-N gauge theory. Recent work on resolving the (AdS-scale) directions transverse to anti-de Sitter space has indeed focused on entanglement between sectors charged under different unbroken gauge symmetries in the Coulomb branch phase of these theories \cite{MollabashiNoburoTadashi:2014,KarchUhlemann}. 

The first significant challenge is isolating degrees of freedom and defining their entanglement with the rest in a gauge-invariant fashion. For the quantum mechanics of a single $N\times N$ Hermitian matrix transforming in the adjoint of the $U(N)$ gauge symmetry, a physical of entanglement and its dual interpretation in 2d string theory was described elegantly in \cite{Das:1995vj, Hartnoll:2015,MazencRenard:2019}. However, in that case, the gauge symmetry leaves only ${\cal O}(N)$ gauge-invariant degrees of freedom, namely the eigenvalues of that matrix up to permutations. Here we will consider matrix quantum mechanical examples with two or more matrices transforming in the adjoint representation, for which there are ${\cal O}(N^2)$ physical, gauge-invariant degrees of freedom.  Through a straightforward gauge fixing we can readily define
a physical notion of ``target space entanglement". Crudely speaking the diagonal elements of the matrices correspond to the positions of
$D$-branes, and the off-diagonal modes to open strings stretching between them. For planar entangling surfaces, we can choose the matrix representing the direction transverse to this surface, use the gauge invariance to diagonalize it, and define an appropriate notion of target space entanglement. This leaves two issues:
\begin{itemize}
    \item This procedure leaves a residual $U(1)^N\rtimes S_N$ gauge symmetry unfixed. The latter is the Weyl group of $U(N)$ and permutes the D-branes.
    \item We must make a physical decision as to the inclusion and treatment of off-diagonal matrix elements corresponding to open strings stretching across the entangling surface.
\end{itemize}

In this note, we attack the above questions by studying bosonic models with two and three matrices. These can be thought of as toy versions of the BFSS model \cite{BFSS:1997}. The residual permutation symmetry is taken care of by considering ``target space entanglement" 
\cite{MazencRenard:2019}, in which we use spacetime position as a quantum number to distinguish particles, much as one would in an EPR experiment involving correlated photons or spin-$\half$ particles. Once this is done, we can think of matrix excitations as living on a complete (all-to-all) lattice, with each vertex corresponding to the location of the diagonal matrix elements, and each link to the strings stretching between these locations. The $U(1)^N$ gauge invariance ensures that the number of strings and anti-strings entering each node have total charge equal to zero. The nodes can be separated according to which side of the entangling surface they reside on. We must then decide how to treat the open strings stretching across the entangling surface. Since we cannot observe one end of these strings, we treat them as unobserved. The resulting reduced density matrix breaks up into superselection sectors according to the observed $U(1)$ charge carried by these strings. This is closely analogous to the discussion in \cite{CHR:2014,TrvediSoni:2015,Donnelly:2012}\ of lattice gauge theories in the electric center definition. Furthermore, for low-energy configurations, the dependence of the state of these strings on the position of the hidden and visible branes induces additional entanglement.

This note is organized as follows: In section \ref{sec:Identical particle entanglement} we review EPR-like entanglement for identical particles. This serves as a primer/review of target space entanglement in the presence of a gauged permutation symmetry, in a simple set of examples. 
%Then, by viewing a system of any number of identical particles as a gauge theory we define a consistent notion of entanglement for a spatial bipartition. 
Next, in section \ref{sec:BFSS} we study toy models with two and three matrices at weak coupling, to understand entanglement between
matrix degrees of freedom.  We draw the connection to entanglement in lattice gauge theories, and point out distinct features involving the off-diagonal degrees of freedom in the two and three-matrix cases which we think captures important qualitative features of entanglement 
in the full BFSS model. 
%We also make connections between such entanglement and lattice gauge theories. 
Finally, in section \ref{sec:future} describe some open problems and directions of future work. Appendix \ref{MatrixMeasure} contains details of the construction of the wavefunction used in the 2-matrix case while appendix \ref{sec:BOCorrections} describes leading order corrections to this wavefunction in the Born-Oppenheimer approximation.

%\textcolor{red} {HRH: Should I include a paragraph with a discussion of numerical studies of bosonic and supersymmetric matrix models and what they have taught us? }There have been several numerical studies of supersymmetric matrix quantum mechanics \cite{Anagnostopoulos:2007fw,Catterall:2008yz,Filev:2015hia,Bergner:2019rca} in the past. Numerical (Monte-Carlo) studies of finite-temperature susy matrix qm model  Monte Carlo studies of bosonic versions \cite{Berkowitz:2016jlq,Azuma:2014cfa} Recent numerical monte carlo explorations \cite{}. Han and Hartnoll \cite{Han:2019wue} have recently used variational Monte-Carlo to study variational wavefunctions of low-energy states in the bosonic three matrix model.

\subsection{Relation to recent work}

While this paper was in preparation, two excellent papers appeared with substantial overlap \cite{Das:2020jhy,Das:2020xoa}. As in our work, the authors study the connection between target space entanglement in Dp-brane field theory and bulk entanglement entropy across surfaces separating Dp-branes: first using the same gauge as our work, \cite{Das:2020jhy}, and then in a gauge-invariant fashion \cite{Das:2020xoa}. These works are complementary and we develop different points, both in the overall explication and in the details of our calculations. To begin with, the focus of that work is on states near to origin of the ``Coulomb branch" in which the eigenvalues of the matrices (aka D0-brane locations) are close together. In this case, given an entangling surface, the wavefunction breaks up into components with different numbers of D0-branes on each side: the reduced density matrix for a spatial region then breaks up into terms with different numbers of D0-branes in that region (as in \cite{MazencRenard:2019}). The coefficients of each of these terms contribute a ``classical" or ``disorder" component to the entanglement entropy. In this work we are interested in well-separated clumps, so we project onto a fixed number.

A second difference is in our treatment of the off-diagonal matrix elements connecting degrees of freedom on each side of the entangling surface. 
In \cite{Das:2020jhy,Das:2020xoa} the authors suggest two different prescriptions to compute entanglement:
\begin{enumerate}
    \item Include off-diagonal matrix elements/ strings stretching across the entangling surface in the subalgebra of observables.
    \item Trace over these strings. 
\end{enumerate}
Our prescription for computing entanglement entropy agrees with 2), and we argue that this is a physical choice. We show that these degrees of freedom lead to the reduced density matrix breaking up into superselection sectors given by the gauge charge carried by strings stretching across the entangling surface, providing a ``classical" component to the entanglement very much in analogy with the ``classical" components that arise in some treatments of lattice gauge theory \cite{CHR:2014,TrvediSoni:2015,Donnelly:2012}. Furthermore, in studying the low-energy states for well-separated branes in the Born-Oppenheimer approximation, we show that the dependence of the state of the off-diagonal modes on the relative separation between the eigenvalues induces further quantum entanglement.

\section{Entanglement for identical particles} \label{sec:Identical particle entanglement}

\subsection{Review: target space entanglement}

In this section we will prepare the way for discussing matrix models by revisiting the question of entanglement for identical particles. The essential reason is that permutation symmetry also will arise in our treatment of matrix models. As a simple example, well discussed in the literature, consider the quantum mechanics of a single $N\times N$ Hermitian matrix which transforms in the adjoint of a $U(N)$ gauge group. We can partially fix the gauge by diagonalizing this matrix and writing a Lagrangian for the eigenvalues. This leaves a residual permutation symmetry which permutes the eigenvalues. This model can then be mapped to a system of non-interacting fermions (see for example \cite{Klebanov:aa,Hartnoll:2015}), in which the residual permutation symmetry of the matrix theory is the permutation symmetry of the identical fermions.

The essential point both for our discussion of matrix models and for other identical particle systems is that the permutation symmetry is a gauge symmetry: if one begins with a Hilbert space that is the product of identical factors, one projects the Hilbert space onto states that are invariant under permutations of these factors up to a sign. The problem is then to find a physically meaningful gauge-invariant notion of entanglement.
To this end, the target space entanglement discussed in \cite{MazencRenard:2019}\ and hinted at in a specific example in \cite{Balachandran-PRL}\  introduces a natural type of entanglement which extends clearly to the matrix models we will study here. We will this focus on this definition.
 
To review, the essential confusion that arises with identical particles is this: Given $N$ particles each described by a one-particle Hilbert space ${\cal H}$, the physical Hilbert space is ${\cal H}^{\otimes N}_{asym}$, where ``asym" denotes a projection 
onto states antisymmetric with respect to the permutation group $S_N$ exchanging factors of ${\cal H}^{\otimes N}$. This
projection amounts to imposing a gauge symmetry. One might have been tempted
to break up the collection into two groups with $k$ and $N-k$ particles and compute a reduced density matrix. 
The antisymmetrization would seem to give a state which is entangled due to the antisymmetrization. But this split is
not gauge invariant as $S_N$ permutes particles between these clumps.

%Hence, the standard construction of a reduced density matrix via a partial trace is not immediately well defined. 
%This leads to the question of how one should define entanglement in these situations. Because there is not a natural factorization one must work hard to find a suitable notion of what it means to split the Hilbert space into two distinct regions. In practice, this is typically accomplished by some choice of projection which takes us to a subspace which does have a natural notion of factorization. In a sense, this is the difficulty of the problem: finding a projection which leads to a meaningful notion of entanglement. For identical particles this is accomplished by making use of the physical separation between them leading to the usual notion of EPR entanglement. 
There are of course clear notions of spatial entanglement for the quantum {\it field} theory of bosonic and fermionic fields, whose excitations correspond to identical particles: for example, one can defines entanglement between regions of the space on which the particles propagate. 
In essence that is what we will do here. However, the matrix model naturally lends itself to a first quantized picture, so we wish to 
understand entanglement in this language. Indeed, in discussions of identical particles, there are circumstances for which a natural first-quantized discussion is desirable. For example, entanglement or lack thereof between two identical spin-$\half$ particles can be described in terms of realizable experiments, of the type attributed to Einstein, Podolsky, and Rosen (more precisely, the version developed by Bohm). This should be describable without quantizing the fermionic field \cite{Dalton_2017}.

To this end, several frameworks have been discussed in the literature to define some notion of entanglement, based on identifying a subclass of observables. Our work will follow from constructions of entanglement based on a sub{\it algebra} of observables. We will follow the specific discussion in \cite{MazencRenard:2019}\ (see also the references in that work) which leads to a notion of entanglement that naturally include EPR-type experiments. A formal definition of entanglement based on identifying operator subalgebras is also given in \cite{Balachandran-PRL,Balchandran:2013a}.

The essential lesson is this: given a subalgebra $A$ of the algebra of observables, one can prove that the Hilbert space decomposes 
into a direct sum of tensor factors,
\begin{equation}
    {\cal H} = \oplus_i {\cal H}_{A,i} \otimes {\cal H}_{{\bar A},i}
\end{equation}
for which the observables $A$ act as the identity operator on the second factor. Given a (pure or mixed) state in ${\cal H}$, we can construct a reduced density matrix that encodes all expectation values of observables in $A$, such that it is a sum over reduced density matrices $\rho_i$ in each subfactor, weighted by a a probability $p_i$. One can the define an entanglement entropy with contributions from a `classical' piece $-\sum_i p_i \ln p_i$ , and a `quantum' piece $S_q = -\sum_i p_i {\rm Tr} \rho_i \ln \rho_i$ which comes from a weighted sum over von Neumann entropies in each subspace. In  \cite{MazencRenard:2019} in particular, the authors describe observables which are constructed as combinations of $\ket{x}\bra{x'}$ where $x,x'$ lie in a specified subregion of the configuration space of the fermions. In the first-quantized language, the index $i$ then corresponds to the number of fermions that sit in this region.

If we further project the state down onto a specific $i$ before computing an entanglement entropy, we clearly describe EPR-type experiments in which each observer is observing a definite number of particles. This is the approach we will take. Since our goal is to describe matrix models which encode D-brane dynamics, it would encode entanglement between clumps of D-branes separated in space, with fixed number of D-brane in each clump. In this setup one {\it can}\ discuss entanglement in an intuitive way, if we understand that we are looking at the entanglement of measurements in two spatial regions. In essence we are using the position basis to ``identify" the identical particles.
Note that this approach, with a discrete ``position" label, was also discussed in an example in \cite{Balachandran-PRL}.

In this section we will review and discuss the target space entanglement of fermions in this way, in the case that space is a finite lattice. We  show how we factorize the Hilbert space for a fixed number of fermions at each position. This is basically working out simple examples of target space entanglement, an exercise we have found useful in laying the groundwork for entanglement in gauged multi-matrix models. 

An alternative treatment of identical particles \cite{Vijay-Entwinement:2018} makes use of an appropriate sub{\it space} spanned by a set of observables, rather than a full subalgebra. This has some antecedents in the quantum chemistry literature. The provides a basis for ``entwinement" in quantum field theories with a gauged permutation symmetry, a quantity which has an appealing holographic dual \cite{Entwinement:2015,Vijay:2016,Vijay-Entwinement:2018}. We leave for the future any application of this approach to gauged multimatrix quantum mechanics.

\subsection{Examples}%EPR-like entanglement} 

The general results discussed above state that the Hilbert space of particles with a permutation symmetry breaks up into a direct sum of factorized 
Hilbert spaces; thus, by projecting onto a summand, we can discuss entanglement in the canonical way. On the way to applying this to the study of gauged matrix models, we have found it useful to work very explicitly through some simple examples involving identical particles on a finite lattice.

\subsubsection{Two spin-$\half$ particles}

%Before describing our framework in detail we would like to motivate our process. To do so we will work 
We begin with the simple case of two indistinguishable spin-$\frac{1}{2}$ particles. Our goal is to provide a concrete example in language
that can be generalized to systems with a larger number of particles. Thus, we will be somewhat pedantic in the hopes that this will pay off
for the reader when we discuss more general examples. 

In the usual EPR-type thought experiment one considers two spin-$\half$ particles which are prepared locally in some state, and then physically separated. The observation that the particles are or are not entangled is then based on first making local measurements of each spin, and then comparing the results. This gives an operational meaning to entanglement between indistinguishable particles: by comparing successful experiments made in each region, we have projected the state of the two particles onto a subspace in which two separate regions each have a definite particle numbers (that is, $1$). In essence we have distinguished the particles by tagging them with their position, and we can construct reduced density matrices in the standard way. 

We next restate this in more precise mathematical language. First, we can get at the essence of physical separation of particles by considering two particles each with spin-$\half$ and with two possible position eigenstates (this model also appears in \cite{Balachandran-PRL}). The position quantum numbers are written as $X = \pm 1 \equiv \pm$. A basis of the four-dimensional Hilbert space of single-particle states is $\ket{\alpha a}$, where $\alpha = \uparrow, \downarrow$ denotes the eigenvalue of $S_z/\hbar$ (with $\uparrow \longrightarrow S_z = \frac{\hbar}{2}$, $\downarrow \longrightarrow S_z = - \frac{\hbar}{2}$), and $a = \pm$ the eigenvalue of the position operator $X$.

Before imposing the permutation symmetry, the Hilbert space of the ``ungauged" two-particle system is 16-dimensional. Projecting onto an antisymmetric wavefunctions leads to a smaller six-dimensional subspace. We can write as a basis $\ket{\alpha a}_1\ket{\beta b}_2 - \ket{\beta b}_1\ket{\alpha a}_2$, although
this includes zero vectors, and so is not the most compact notation. On the other hand, we can separate the spin and position degrees of freedom, so that beginning with basis vectors $\ket{ab}\ket{\alpha\beta} \equiv \ket{\alpha a}_1\ket{\beta b}_2$, the antisymmetric states are:
%It is instructive to construct the antisymmetric wavefunctions that span this space. These can be built by taking products of symmetric wavefunctions of the two particles in a given basis (spin/isospin) and antisymmetric wavefunctions in the other to make the total wavefunction antisymmetric.
%
\begin{align} \label{antisymm_2Dwavef}
\begin{split}
\psi_1 & =\frac{1}{\sqrt{2}}|+ + \rangle \otimes \big(|\uparrow \downarrow \rangle - |\downarrow \uparrow \rangle\big) \\
\psi_2 & =\frac{1}{\sqrt{2}}|-- \rangle \otimes \big(|\uparrow \downarrow \rangle - |\downarrow \uparrow \rangle\big)  \\
\psi_3 & =\frac{1}{2}\big(|+ - \rangle  + | -+ \rangle \big) \otimes \big(|\uparrow \downarrow \rangle - |\downarrow \uparrow \rangle \big) \\
\psi_4 & =\frac{1}{\sqrt{2}}\big(|+ - \rangle  - | -+ \rangle \big) \otimes |\uparrow \uparrow \rangle  \\
\psi_5 & =\frac{1}{\sqrt{2}}\big(|+ - \rangle  - | -+ \rangle \big) \otimes |\downarrow \downarrow \rangle  \\
\psi_6 & = \frac{1}{2}\big(|+ - \rangle  - | -+ \rangle \big) \otimes \big (|\uparrow \downarrow \rangle + |\downarrow \uparrow \rangle\big) \\
\end{split}
\end{align}
Let us consider further the subspace of states for which the two particles are physically separated, which is the image of the projection $ P =  P^{pos} \otimes 1^{spin} $ where  $P^{pos} = |+ - \rangle \langle +-| + |-+ \rangle \langle -+|$. This leaves a four-dimensional Hilbert space spanned by $\psi_{3,4,5,6}$. 

%These are the six linearly independent antisymmetric wavefunctions which are nothing, but the product of appropriate EPR pairs in each basis. The essence of EPR entanglement is captured by distinguishing the two particles by using their isopin labels before computing the von Neumann entropy in the spin basis. This is accomplished by using the projector $ P = 1^{spin} \otimes P^{iso}$ where  $P^{iso} = |+ - \rangle \langle +-| + |-+ \rangle \langle -+|$. This projects out all states where the particles are not separated: the first two wavefunctions in \eref{antisymm_2Dwavef}. We are left with a four-dimensional Hilbert space spanned by four EPR/Bell pairs.

At this point we could simply fix the gauged permutation symmetry by placing particle 1 at position $-$ and particle 2 at position $+$. Nonetheless, we find it illuminating to provide a more manifestly $S_2$-invariant setup. Working with the basis $\psi_{3,4,5,6}$, we seek a natural factorization $\IC^4 = \IC^2\otimes\IC^2$ corresponding to the usual EPR notion of entanglement in which we measure spins that are associated to specific position quantum numbers. To this end, we introduce the operators
\begin{align} \label{projection_and_measurement_op}
\begin{split}
\mathbb{O}_{+ z}  & := 1 \otimes P_{+} \otimes 1 \otimes \sigma_z +   P_{+} \otimes 1 \otimes \sigma_z \otimes 1 \\
\mathbb{O}_{- z} & := 1 \otimes P_{-}\otimes 1 \otimes \sigma_z +   P_{-} \otimes 1 \otimes \sigma_z \otimes 1 
\end{split}
\end{align}
where we have written these in the same basis as (\ref{antisymm_2Dwavef}), in which the first two factors label the position and the second two factors label  the spin. These operators measure the spins associated to specific position states.
They are diagonal in the basis spanned by:
 \begin{equation}
 \begin{split}
 |1 \rangle :=  \frac{1}{\sqrt{2}}(\psi_3 + \psi_6), \quad |2 \rangle :=  \frac{1}{\sqrt{2}}(\psi_3 - \psi_6)\\
 |3 \rangle :=  \frac{1}{\sqrt{2}}\psi_4, \quad |4 \rangle :=  \psi_5
 \end{split}
 \end{equation} 
in which they can be written as  $\IO_{+z} = diag(1,-1,1,-1)$ and $diag(-1,1,1,-1)$ respectively. One can check quickly that the operators do not have support outside this subspace: they annihilate $\psi_{1,2}$. Within this subspace, $\IO_{\pm,z}$ are each doubly degenerate but together form a complete set of commuting observables: their eigenstates thus naturally describe $\IC^4$ as $\IC^2\otimes\IC^2$. 
It is straightforward to check that the states $\ket{k}$ are the four states 
\be \label{final_projected_states}
 \frac{1}{\sqrt{2}}(| + - \rangle |ab \rangle - |- + \rangle |ba \rangle) , \quad a,b= \uparrow, \downarrow .
\ee 
in which each position can be associated to a definite value of the spin. From these we can take linear combinations to construct nontrivial EPR pairs. We can simplify the notation further by labeling the states as $\ket{\IO_{+,z},\IO_{-,z}}$ according to the eigenvalues of $\IO_{\pm,z}$. Here the first label is the spin at the position $+$ and the second the spin at the position $-$. In this language,
 \begin{equation}
 \begin{split}
 |1 \rangle &:= |1,-1 \rangle \\
 |2 \rangle &:= |-1,1 \rangle \\
 |3 \rangle &:= |1,1 \rangle \\
 |4 \rangle &:= |-1,-1 \rangle \\
 \end{split}
 \end{equation} 
and the basis is that of two distinguishable spin-$\half$ particles in $\IC^2\otimes\IC^2$, for which standard notions of entanglement can be defined. In particular
linear combinations such as 
\begin{eqnarray}
    \frac{1}{\sqrt{2}} \big(\ket{1} - \ket{2}\big) & = & \psi_6\nonumber\\
    & = & \frac{1}{\sqrt{2}}\big(\ket{1,-1} - \ket{-1,1}\big)
\end{eqnarray}
describe the entanglement of separated particle in a way that is gauge-invariant and intuitive. It is clear, for example, that observations with $\IO_{+,z}$ alone are described by a mixed state with density matrix equal to the identity operator on $\IC^2$; this operator can be computed via a standard partial trace.

%Tracing over $\mathbb{H}_{2}$, we find that $ \rho_{1} =  \frac12 \Big(|1 \rangle \langle 1 | + |-1 \rangle \langle -1 )\Big)$ which has a von Neumann entropy $S= \log2$. We have successfully described the entanglement between an EPR pair in a gauge-invariant fashion. Note that this is not a new idea and is morally very similar to the method of \cite{Balachandran-PRL,Balchandran:2013a}, however they get the EPR entanglement by restricting to a proper subalgebra of the algebra of one-particle observables \footnote{Note that we cannot use the subalgebra of all the one-particle observables as that would end up being equal to algebra of all observables in the system. Alternatively, \cite{Vijay-Entwinement:2018} has suggested that we should use a subspace (of the full algebra of all observables) spanned by one-particle observables.} by working only with the spin degrees of freedom and pretending that the particles are distinguishable. Their proper subalgebra involves operators that are not appropriately symmetrized and thus, are not gauge invariant. Thus, one can make sense of entanglement in identical particles by first `distinguishing' them using some other label and working in the projected Hilbert space which factors into two subspaces in the appropriate basis. 

\subsubsection{$3$ spin-$\half$ particles}

As a second example, let us work out the example 3 spin-$\half$ particles on a three-site lattice. 
The position basis is $\ket{X=1,2,3}$; for spin we again take the basis  $\ket{m = \pm \half}$ of eigenstates of $\sigma_z$. The full Hilbert space is 
\be 
    {\cal H} = \left(\IC^3_{position}\otimes\IC^2_{spin}\right)^{\otimes 3}_{asymm}
\ee
Following the discussion above of ``target space entanglement", we wish to consider the entanglement of a particle at $X = 1$ with two other particles at $X \in \{2,3\}$. The corresponding subalgebra of observables is that generated by 
\be
    {\cal O}_{1\mu} = \left({\IP}_1\otimes\sigma_\mu\right)\otimes {\bf 1}\otimes{\bf 1} + 
    {\bf 1}\otimes \left({\IP}_1\otimes\sigma_{\mu}\right)\otimes {\bf 1} + 
    {\bf 1}\otimes {\bf 1}\otimes \left({\IP}_1\otimes\sigma_{\mu}\right)
\ee
where $\mu \in \{0,1,2,3\}$, $\sigma_0$ is the identity operator on $\IC^2$, $\sigma_{1,2,3}$ are the Pauli matrices acting on $\IC^2$, and $P_1 = \ket{1}\bra{1}$ is the projector onto $X = 1$ in the Hilbert space of positions. The operators ${\cal O}_{1\mu}$ generate $1-$ and $2-$ particle observables for fermions located at $X = 1$. The Hilbert space ${\cal H}$ can be repackaged as
\be
    {\cal H} = {\cal H}^{(3)}_{X = 2,3} \oplus \left({\cal H}_{X=1}^{(1)} \otimes {\cal H}_{X=2,3}^{(2)}\right) \oplus \left({\cal H}^{(2)}_{X=1}\otimes {\cal H}^{(1)}_{X=2,3}\right)
\ee
where ${\cal H}^{(k)}_{X=i_1\ldots i_p}$ corresponds to the $k$-particle Hilbert spaces for particles restricted to the sites $i_1\ldots i_p$.
This factorization is manifest in the occupation number basis, in which we label the states by the number of particles for each value of $X$ and $\sigma$, with the total number of particles set to $N = 3$ and the Pauli exclusion principle imposed. In this basis, it is convenient to work in second quantized language, with fermionic annihilation and creation operators $a_{i\alpha},a_{i\alpha}^{\dagger}$, $i \ 1\ldots 3$, $\alpha = \pm \half$. A complete basis of states consists of all non-vanishing combinations %
\be
    \ket{i_1\alpha_1;i_2\alpha_2;i_3\alpha_3} = a^{\dagger}_{i_1,\alpha_1}a^{\dagger}_{i_2,\alpha_2} a^{\dagger}_{i_3,\alpha_3}\ket{0}
\ee
where $\ket{0}$ is the fermionic vacuum. Note that in this language,
\be
    {\cal O}_{1\mu} = a_{1\alpha}^{\dagger}\sigma_{\mu,\alpha\beta}a_{1\beta}. 
\ee
The basis states $\ket{i_1\alpha_1;i_2\alpha_2;i_3\alpha_3}$ will have zero entanglement between particles localized at $X = 1$ and particles located at $X = 2,3$: entangled states are linear combinations of these basis states.

The unentangled basis $\ket{i_1\alpha_1;i_2\alpha_2;i_3\alpha_3}$ can also be written in position space as:
\be
    \ket{i_1\alpha_1;i_2\alpha_2;i_3\alpha_3} \equiv \frac{1}{\sqrt{6}}\sum_{\sigma \in S_3} (-1)^\sigma
        \ket{X_{\sigma(1)}\alpha_{\sigma(1)}}
        \ket{X_{\sigma(2)}\alpha_{\sigma(2)}}
        \ket{X_{\sigma(3)}\alpha_{\sigma(3)}}
        \label{eq:posspacebasis}
\ee
This latter presentation is closer to the presentation that  naturally emerges from considering gauged matrix models. Note that if we were to embed this state into a state of three distinguishable particles in the Hilbert space $\left(\IC^3_{position}\otimes\IC^2_{spin}\right)^{\otimes 3}$, and compute the entanglement of the particle in the first factor with the others,  we would find that the first particle is entangled with the other two. This is not target space entanglement: for each term in the sum the particle in the first factor of $\left(\IC^3_{position}\otimes\IC^2_{spin}\right)^{\otimes 3}$  is at a distinct location. More importantly, it is not gauge invariant (where we take the action of $S_3$ as a gauge symmetry). A gauge-invariant quantity is the target space entanglement of a single particle at $X = 1$ with two particles at $X \in \{2,3\}$. We could compute this within the space $\left(\IC^3_{position}\otimes\IC^2_{spin}\right)^{\otimes 3} = \IC^{15}$ 
by embedding the $\sigma = {\bf 1}$ term in  (\ref{eq:posspacebasis}), considering linear combinations of such states, and computing the entanglement between the first particle and the other two. This amounts to a gauge fixing with respect to $S_{k=3}$, to a gauge in which the
unentangled nature of the basis vectors (\ref{eq:posspacebasis}) is clear. In practice this is what we will do below, but we emphasize that we could phrase our entanglement in such a way that the $S_k$ gauge symmetry is manifest.

In physical systems energy eigenstates of $N$ particles may, depending on the Hamiltonian, be more naturally constructed from states with definite properties under $S_N$ for the position and spin degrees of freedom separately. The symmetry types are coupled by the demand that the total wavefunction be antisymmetric. To see how this works out, we follow the textbook construction of these states via Young diagrams. (See for example \cite{hamermesh2012group} for a clear discussion). We will see that such wavefunctions are entangled in the target space.

Recall that an irrep $R$ of $S_N$ is determined by a Young diagram with $N$ boxes. From each of these diagrams one may construct a Young operator $Y$ acting as a projection operator 
\be
    Y: {\cal H}^{\otimes N} \longrightarrow {\cal H}_R
\ee
Given a basis $\ket{i = 1,\ldots n}$ of ${\cal H}$, the elements of  
${\cal H}_R$ correspond to $N$-rank tensors of fixed symmetry type:
\be
    T_{i_i\ldots i_N} \ket{i_1}\ldots\ket{i_N}
\ee
To construct a basis in ${\cal H}_R$, we choose $k$ numbers from $\{1,\ldots n\}$ and fill in the Young diagram to create a standard tableaux so that the numbers are nondecreasing along rows and strictly increasing along columns. Reading these numbers from left to right along the top row, then the second row, etc, we assign them in turn to $i_1\ldots i_n$; we then act on $\ket{i_1}\ldots\ket{i_N}$ with the associated Young operator. 

In our examples with two quantum numbers, the position and spin wavefunctions must be constructed from conjugate Young diagrams, with the rows and columns transposed, for the total wavefunction to be antisymmetric. The specific formula for combining the wavefunctions can be found, for example, in \cite{hamermesh2012group} (section 7-14). 

%
%Given $n$ labels that correspond to some degree of freedom (like spin), we assign numbers to each boxes so that that numbers do not decrease along a row (symmetric combinations) and must strictly increase along a column (antisymmetric combinations). A diagram which is just a row of $d$ boxes corresponds to totally symmetric \textit{wavefunctions} and a pure column corresponds to totally antisymmetric wavefunction. With these rules, it is clear that a totally antisymmetric wavefunction for 3 spin-$\frac12$ particles is impossible.}  There is one more input that we need- in order for the total wavefunction to be asymmetric, the Young diagrams of the spin (denoted by $\chi$) and coordinate/isospin (denoted by $\phi$) wavefunctions must be dual in relation i.e, they must be obtained by exchanging rows and columns. Armed with these rules, let us calculate the wavefunctions:

Let us consider the case that the positions of the particles are distinct, and the individual particles have spin-$\half$.
The Young diagrams, and the number of states for each quantum number, are shown below:
\begin{align} \label{youngdiag_3_2}
\begin{split}
spin: \text{ (box values 1,2) }  \chi : \yng(3) \quad (4\; states) &\qquad \yng(2,1) \quad (2\; states) \\
position: \text{ (box values 1,2,3) }  \phi : \qquad  \yng(1,1,1) \quad (1\; state ) &\qquad \yng(2,1) \quad (2\; states )
\end{split}
\end{align}
Here the counting is understood by the number of ways of filling up the boxes with $\alpha = 1,2$ (corresponding to spins $m = -\half,\half$) for spin or $n = 1,2,3$ (corresponding to ${\vec x}_1,{\vec x}_2,{\vec x}_3$) for positions, so that the numbers are non-decreasing along rows and increasing along columns. For example, in the first line of \ref{youngdiag_3_2}, the boxes corresponding to completely symmetric spin wave function can be filled up in four different ways $ (1,1,1), (1,1,2), (1,2,2) , (2,2,2)$ where $1$ refers to spin up and $2$ refers to spin down. \\

Let us again focus on states for which each particle is at a distinct position. The completely asymmetric state in line 2 of \ref{youngdiag_3_2} survives and gives rise to four wavefunctions. Among the eight states of mixed symmetry, only two - (1,2,3) and (1,3,2)- survive and contribute a total of four wave functions so that the dimension of the resulting Hilbert space is $8 = 2^3$. We can use the basis described in the previous section to write these wavefunctions; this is a nontrivial change of basis, so that the states constructed as above from Young tableaux are naturally entangled from the point of view of the target space. We first define the completely antisymmetric position wavefunction and a partially antisymmetrized wavefunction respectively as follows:
\begin{align}
\begin{split}
| r_1,r_2,r_3 \rangle_{as} &= \frac{1}{\sqrt{6}} \left(|r_1,r_2,r_3 \rangle - |r_2,r_1,r_3 \rangle + |r_3,r_1,r_2 \rangle - |r_3,r_2,r_1 \rangle\right.\\
& \qquad \qquad \qquad \left. - |r_1,r_3,r_2 \rangle + |r_2,r_1,r_3 \rangle\right) \\
|\chi_{123}; \frac12 \rangle_{as}  & =|\uparrow \rangle_{1} \frac{1}{\sqrt{2}} \big(| \downarrow \rangle_{2} |\uparrow \rangle_{3} - |\uparrow \rangle_{2} |\downarrow \rangle_{3} \big) \\
|\chi_{123}; -\frac12 \rangle_{as}  & =|\downarrow \rangle_{1} \frac{1}{\sqrt{2}} \big(| \downarrow \rangle_{2} |\uparrow \rangle_{3} - |\uparrow \rangle_{2} |\downarrow \rangle_{3} \big).\\
|\chi_{123}; \frac12 \rangle_{s}  & =|\uparrow \rangle_{1} \frac{1}{\sqrt{2}} \big(| \downarrow \rangle_{2} |\uparrow \rangle_{3} + |\uparrow \rangle_{2} |\downarrow \rangle_{3} \big) \\
|\chi_{123}; -\frac12 \rangle_{s}  & =|\downarrow \rangle_{1} \frac{1}{\sqrt{2}} \big(| \downarrow \rangle_{2} |\uparrow \rangle_{3} + |\uparrow \rangle_{2} |\downarrow \rangle_{3} \big).
\end{split}
\end{align}
where we the position labels take the values $r_1,r_2$ and $r_3$. Similarly, we can also define the position eigenfunctions symmetric and antisymmetric respectively w.r.t exchange of particles $2$ and $3$,
\begin{eqnarray}
|\phi_{123} \rangle_{s} & = & |r_1 \rangle \frac{1}{\sqrt{2}} \big( |r_2 \rangle |r_3 \rangle + |r_3 \rangle |r_2 \rangle \big)\nonumber\\
|\phi_{123} \rangle_{as} & = & |r_1 \rangle \frac{1}{\sqrt{2}} \big( |r_2 \rangle |r_3 \rangle - |r_3 \rangle |r_2 \rangle \big).
\end{eqnarray}

Using these basis states for the position and spin wavefunctions we can write a complete basis for the full Hilbert space of spin-$\half$ particles in distinct positions:
\begin{align} \label{antisymm_3Dwavef}
\begin{split}
\psi_1 & =|\uparrow \uparrow \uparrow \rangle \otimes | r_1,r_2,r_3 \rangle_{as} \\
\psi_2 & =\frac{1}{\sqrt{3}} \big(|\uparrow \uparrow  \downarrow \rangle + |\uparrow \downarrow \uparrow \rangle + |\downarrow  \uparrow \uparrow  \rangle \big) \otimes | r_1,r_2,r_3\rangle_{as}  \\
\psi_3 & =\frac{1}{\sqrt{3}} \big(|\uparrow \downarrow  \downarrow \rangle + |\downarrow \uparrow \downarrow \rangle + |\downarrow \downarrow \uparrow  \rangle \big) \otimes |r_1,r_2,r_3\rangle_{as} \\
\psi_4 & = |\downarrow \downarrow \downarrow \rangle \otimes |r_1,r_2,r_3 \rangle_{as} \\
\psi_5 & = |\chi_{123}; \frac12 \rangle_{as} \otimes  |\phi_{123} \rangle_{s} + |\chi_{231}; \frac12 \rangle_{as} \otimes  |\phi_{231} \rangle_{s} + |\chi_{312}; \frac12 \rangle_{as} \otimes  |\phi_{312} \rangle_{s}  \\
\psi_6 & =  |\chi_{123}; -\frac12 \rangle_{as} \otimes  |\phi_{123} \rangle_{s} + |\chi_{231}; -\frac12 \rangle_{as} \otimes  |\phi_{231} \rangle_{s} + |\chi_{312}; -\frac12  \rangle_{as} \otimes  |\phi_{312} \rangle_{s}\\
\psi_7 & = |\chi_{132}; \frac12 \rangle_{s} \otimes  |\phi_{132} \rangle_{as} + |\chi_{213}; \frac12  \rangle_{s} \otimes  |\phi_{213} \rangle_{as}
+ |\chi_{321}; \frac12  \rangle_{s} \otimes  |\phi_{321} \rangle_{as}  \\
\psi_8 & =  |\chi_{123}; -\frac12 \rangle_{s} \otimes  |\phi_{123} \rangle_{as} + |\chi_{231}; -\frac12 \rangle_{s} \otimes  |\phi_{231} \rangle_{as} + |\chi_{312}; -\frac12  \rangle_{s} \otimes  |\phi_{312} \rangle_{as}\\
\end{split}
\end{align}

Let us consider the state $\psi_3$. In the occupation number basis discussed above this corresponds to the state
\be
    \frac{1}{\sqrt{3}}\left[\ket{x_1,-\half;x_2,-\half,x_3,\half} \ket{x_1,-\half;x_2,\half,x_3,-\half}
    + \ket{x_1,\half;x_2,-\half,x_3,-\half}\right]
\ee
The corresponding reduced density matrix for the particle at $x_1$ is:
\be
    \rho = \frac{2}{3}\ket{-\half}\bra{-\half} +
    \frac{1}{3}\ket{\half}\bra{\half}
\ee
That is, there is nontrivial target space entanglement between a particle at any fixed position, and the two particles at other positions.

We could continue in this vein with more particles, and more generally partitioning the ``target space" between two groups, with fixed numbers of particles in each group; these particles may or may not be coincident in position within either group. The upshot will be the same: the occupation number basis will naturally yield states that are unentangled from the standpoint of target space entanglement, while the states constructed from irreps of the permutation group acting on each of the quantum numbers will naturally yield entangled states.

\section{Entanglement in gauged matrix models}\label{sec:BFSS}

In this section, we set up a framework to compute target space entanglement in low-energy states of simple bosonic gauged matrix models of  two and three matrices. We view this as a step to understanding entanglement in large-N theories relevant for holography, for example
those which arise as the strong-coupling, low-energy limits of D-brane dynamics. The gauge theories under anything like computational control on either side of the holographic duality are typically supersymmetric, contain more bosonic matrices, and often extended spatial directions. 
In particular, supersymmetry allows for low-energy states of well-separated D-branes, and as such ensures the consistency of the Born-Oppenheimer approximation we use to describe these configurations.  For the sake of clarity, we will work with the reduced bosonic models, and assume the cancellation of the ground state energy occurs. At the lowest order in which we would require this effect, the extra degrees of freedom will simply contribute additively to the entanglement  .

%
% however, a generalization to include more bosonic and fermionic matrices should be straightforward as we explain below. Our motivation is to understand the role of entanglement in theories such as the BFSS matrix model \cite{BFSS:1997}, where 11-d spacetime emerges from the strong coupling limit of a quantum mechanical theory of D0 branes and $\mathcal{N}=4$ SYM. However, despite the lack of realism by restricting to simple models, we are able to describe important features which we believe should carry over to these more physically motivated theories.
%We also point out how calculating entanglement in states with well-separated branes can be mapped to the computation of entanglement in Hamiltonian lattice gauge theories.

\subsection{Outline of an entanglement entropy calculation}

In order to motivate the specific form of our simplified models, we start by recalling the action \cite{BFSS:1997}\ for the
BFSS matrix model, which can be attained by dimensional reduction (in flat space) of the 10-d maximally supersymmetric 
SYM action to 0+1 -dimensions:
\be \label{BFSSAction}
S =  \frac{1}{e^2 l_s }\int dt \, \text{Tr} \Bigg[ \frac12 \sum_{i=1}^{9} D_t X^2_{(i)} - \frac{1}{l_s^4} \sum_{i\neq j} \ [X_{(i)},X_{(j)}]^2 + \text{fermions} \Bigg].
\ee 
$X_{(i)} $ are nine bosonic Hermitian $N \times N$ matrices with dimensions of length, and $ D_t{X}_{(i)} = \del_t X_{(i)} + [A_{(0)}, X_{(i)}]$, 
where $A_0$ is a $0+1$-dimensional $U(N)$ gauge field.  This describes the massless sector of open strings ending on $N$ D0-branes in $9+1$ flat spacetime dimensions, with spatial coordinates on this spacetime as $Y^{i = 1,\cdots 9}$, and with target space and worldline time identified. It is well known that at sufficiently low energies, closed strings as well as oscillator modes of open strings decouple.

The theory has a $U(N)$ gauge symmetry, with all fields transforming in the adjoint of the gauge symmetry.  The index $(i)$ in parentheses is a ``flavor" index, identified with the index for target space coordinates. Thus the bosonic fields transform as $SO(9)$ vectors. We will use unbracketed subscripts $a, b = 1\ldots N$ to refer to the matrix elements, on which the gauge symmetry acts. Thus in index notation we can write the bosonic fields as $X_{(i)ab}$. The 0+1- dimensional gauge coupling $e^2=g_s$, where $g_s$ is the dimensionless string coupling. The natural dynamical length scale in the problem is $l_p^{11}= (g_s)^{\frac13}l_s$ 
\cite{Danielsson:1996uw,Kabat:1996cu}, the eleven-dimensional Planck scale; this denotes the size of bound states at threshold.

We will need to pay careful attention to two features of this theory. First, the gauge symmetry must be treated properly. For the specific questions we will answer, we will do this by choosing a gauge appropriate to answering those questions, and imposing the residual gauge invariance on quantum states. Secondly, the fermionic variables in the theory are crucial to understanding D0-brane dynamics. In particular, they ensure the existence of low-energy states of well-separated branes. That said we will for the most part ignore them, and focus our studies on entanglement in reduced bosonic models.
we believe the lessons we draw from the bosonic sector should generalize when fermions are included. While fermions give a negative contribution to the vacuum energy, they generally contribute positively to entanglement. 

%Motivated by the previous studies of entanglement in the Coulomb branch in $\mathcal{N}=4, d=4$ SYM \cite{MollabashiNoburoTadashi:2014,KarchUhlemann}, 
We imagine low-energy states which are well localized about configurations far along the flat directions of the potential in \eref{BFSSAction}, corresponding to well-separated D-branes. The separation of the D0-branes amounts to a Higgsing of the gauge symmetry $U(N) \rightarrow U(I)^N$. It is well known that in this `classical' regime, the diagonal entries of these matrices can be interpreted as the coordinates of the D0-branes in the 9+1-dimensional bulk theory. The off-diagonal variables represent the fields for strings stretching between the D0-branes.  Fermionic variables will cancel the zero-point energy of the off-diagonal strings, and ensure there is no static potential between the D0-branes.

We must be careful about gauge invariance, but this identification is close to the gauge-invariant reality if the branes are well-separated.  We could consider explicitly gauge-invariant variables by writing $X_{(i)} = U_{(i)}^{\dagger} \Lambda_{(i)} U_{(i)}$, where $U$ is a unitary matrix sometimes denoted the ``angular components" and $\Lambda_{(i)}$ is the real, diagonal matrix of eigenvalues. In the limit of well-separated branes, the degrees of freedom denoted by $U_{(i)}$ become massive with masses equal to the difference between eigenvalues; the eigenvalues of $\Lambda_{(i)}$ are gauge-invariant up to discrete permutations, and can be approximately identified with the diagonal elements $X_{ii}$. The discrete permutations reflect the fact that the D0 branes should be treated as identical particles. 

We will work with a specific target space entanglement calculation that lends itself to a simple gauge fixing. We consider states in which $n$ D0-branes are localized in the region $Y^1 < 0$ and the remaining $N - n$ localized in the region $Y^1 > 0$. We can effectively use the position of the branes to label them, as in \S 2. We can then ask the question: `how entangled are the branes and strings within the region $X_{(1) ii} <0$ with the strings and branes in the region $X_{(1) 11} >0$?'. 

We begin by fixing to the axial gauge $A_t = 0$. Once done, one still has to impose the condition that $\delta L/\delta A_t = 0$; this condition is identical to the demand that the wavefunction $\Psi(X^{(i)})$ is invariant under $U(N)$ transformations,
\be
\Psi(X_{(i)}) = \Psi(U^{\dagger} X_{(i)} U)\label{eq:wftransform}
\ee
where $U$ is a $U(N)$ matrix: to see this, consider infinitesimal transformations in \eref{eq:wftransform}. For the question at hand, it makes sense to use this invariance to diagonalize $X_{(1)}$: the resulting components of the matrix are precisely the eigenvalues of that matrix in the full theory.  Note that this does not completely fix the remaining gauge invariance, as this choice for $X_{(1)}$ is unchanged by a residual $U(1)^N\rtimes S_N$ when the eigenvalues are distinct. For simplicity we will focus on this case. Under the residual gauge group, the matrix elements $X_{(i > 1)ab}$ have charge $(-1,1)$ under $U(1)_a \times U(1)_b$. Physical states in this theory will have charge zero under this residual $U(1)^N$. So long as our entangling surface is planar (and the target space flat), our gauge choice is natural.

As is well known, this gauge fixing of $X_{(1)}$ means that the measure of the path integral includes a Vandermonde determinant for the eigenvalues of $X_{(1)}$. This determinant can be included in the wavefunction so that the measure on the eigenvalues is flat. The result is that under actions of the $S_N$ Weyl group, the wavefunction should transform with a sign equal to the sign of the permutation.
(See appendix \ref{MatrixMeasure} as well as \cite{Martinec:2004td,Das:2020jhy}.) If the eigenvalues of the matrices are distinct, we can fix the permutation symmetry by ordering the eigenvalues so that $X_{11} < X_{22} < \ldots X_{NN}$.

Our desire is to compute the entanglement entropies (of von Neumann or R\'enyi type) between degrees of freedom localized on each side of $Y^1 = 0$. The negative elements of $X^{(1)}$ define an $n\times n$ block (after relabeling the indices), corresponding to strings stretching between the visible branes. We consider the entanglement of all of the matrix elements of $X^{(i)}$ within this block with the remaining degrees of freedom.

In doing so, there are off-diagonal matrix elements which couple the visible and ``hidden" blocks of the matrix fields. These correspond to string fields for open strings stretching across this entangling surface. We must decide how to treat them. In string language, these strings are labeled by which D-branes they begin and end on. In the full string theory, the open strings would have internal degrees of freedom controlling their vibrational and rotational states, about which at least partial information could be gained via closed string scattering experiments done in the $Y^{(1)} < 0$ region.\footnote{Such a calculation could be done via string field theory \cite{Balasubramanian:2018axm}.} In the low-energy limit we consider, such  information is invisible to us. What we can record is the total $U(1)^n$ charge carried by strings stretching across the entangling surface. This must be cancelled by additional strings stretching between the D0-branes localized in the $Y^{(1)} < 0$ region, in order to satisfy the Gauss' law constraint that the total gauge charge be zero. As we will discuss in detail below, the full Hilbert space will break up into superselection sectors labeled by this $U(1)^n$ charge. It is possible that one could get additional information about these string fields and their coupling to the other hidden degrees of freedom, by measuring the force they exert on the D-branes at $Y^{(1)} < 0$. We leave this possibility aside for now. The upshot, as we will describe in more detail in \S 3, is that the density matrix for the visible degrees of freedom will break up into pieces with different external $U(1)^n$ charge, leading to a ``classical" contribution to the entanglement entropies entirely which is analogous to the classical contribution that appears in the Extended Hilbert Space formulation of entanglement in lattice gauge theory \cite{CHR:2014,TrvediSoni:2015,Donnelly:2012}. 

While we have discussed our problem in terms of target space entanglement of D0-branes, which gives a clear geometric picture, we can 
phrase the story more directly in terms of the gauge theory. Starting with a $U(N)$ gauge theory with adjoint matter, we consider configurations which spontaneously break the gauge symmetry according to the pattern
\be
U(N) \longrightarrow U(n) \times U(N - n) \rightarrow U(1)^n \times U(1)^{N - n} = U(1)^N.
\ee
We are interested in the entanglement of the degrees of freedom that transform as adjoints of the $U(n)$ factor. The density matrix
will break up into superselection sectors based on the total $U(1)^n$ charge carried by degrees of freedom which are ``bifundamental"
under the group $G_+ \times G_-$ where $G_- = U(1)^n$, $G_+ = U(1)^{N-n}$.

We will now proceed to a more explicit discussion in terms of reduced models.

\subsection{Entanglement in a two-matrix model } \label{TwoMatrixEntanglemnent}

\def\ket#1{{| #1 \rangle}}
\def\bra#1{{\langle #1 |}}

We first consider a bosonic theory with just two Hermitian $N\times N$ matrices, given by the action,
\be \label{TwoMatrixAction}
S = \frac{1}{2e^2l_s} \int dt \, \text{Tr} \Bigg( D_t X^2+ D_t Y^2 -\frac{1}{l_s^4} [X,Y]^2 \Bigg).
\ee
where $D_t$ is the gauge covariant derivative for a $U(N)$ gauge group which acts on the bosonic Hermitian matrices $X,Y$ in the adjoint representation. This is a crude model for $N$ D0-branes in 2 spatial dimensions. Up to gauge transformations, the degrees of freedom of this theory are the eigenvalues and angular directions of these matrices. Following the discussions in \S2\ and \S3.1, we wish to compute entanglement entropies for a given state after projecting onto the subspace for which $k < N$ eigenvalues of $X$ have negative values, corresponding to $k$ D-branes localized at $Y^{(1)} < 0$.

As above, we choose axial gauge and further solve for the remaining gauge invariance by gauging away the 
angular directions of $X$. We restrict our states by demanding they are invariant under the residual $U(1)^N$ gauge invariance,
and sum over the images of this state under actions of the Weyl group. For a given image, 
the $k$ negative diagonal components $X$ define a block; we then construct the reduced density matrix for all degrees of freedom of $X$, $Y$ that live within that block, breaking it up into superselection sectors corresponding to the $U(1)^k$ charge of the strings that
are bifundamental with respect to the decomposition $G_k \times G_{N-k}$ with $G_n = U(1)^n$.

We will be interested in constructing the reduced density matrix for states that are models of low-energy states of D0-branes.
To this end we proceed by writing out the Hamiltonian more explicitly.
%We will find it 
%As we proceed we will be careful about our choice of gauge and any residual gauge symmetry that is left. To begin, let us diagonalize the matrix $X$ using the $SU(N)$ gauge symmetry so that $X= UDU^{\dagger}$. The non-zero elements of the resulting diagonal matrix are the $X$-coordinates of the D0-branes. We will choose our entangling surface to be perpendicular to the $X$-direction. 
We will use lower case letters to denote the diagonal matrix elements of $X$ and $Y$, and denote them by a single index. For example, $(X)_{11}:=x_{11}=x_1$. The Hamiltonian corresponding to \eref{TwoMatrixAction} can now be written as:
%Apart from the $U(1)^N$ symmetry, there is a residual permutation symmetry $S_N$ of the eigenvalues of the matrices. 
\begin{align} \label{BO-Hamiltonian}
	\begin{split}
		H            &= H_{slow} + H_{fast}, \\
		H_{slow} & = H_{diag} = \sum _{i=1}^{N}\frac{e^2 l_s}{2}  (\pi_{x_i}^2 +\pi_{y_i}^2) , \\
		H_{fast} & = H_{off-diag} = \sum_{i<j,i=1}^{N} \Bigg(e^2 l_s\pi^{\dagger}_{Y_{ij}}\pi_{Y_{ij}} + \frac{1}{e^2 l_s^5} (x_i-x_j)^2 |Y_{ij}|^2 \Bigg), 
	\end{split}
\end{align}
where we have used the Hermiticity of $Y$ to equate $Y_{ij} = Y_{ji}^{\ast}$, $\pi_{Y_{ij}} = \pi^{\ast}_{Y_{ji}}$.
%\begin{align} \label{BO-Hamiltonian}
%\begin{split}
%H            &= H_{slow} + H_{fast}, \\
%H_{slow} & = H_{diag} = \frac{e^2 l_s}{2}  (\pi_{x_1}^2 +\pi_{x_2}^2 + \pi_{y_1}^2 +\pi_{y_2}^2), \\
%H_{fast} & = H_{off-diag} = \frac{e^2 l_s}{2} (\pi^{\dagger}_{Y_{12}}\pi_{Y_{12}}+ \pi^{\dagger}_{Y_{21}} \pi_{Y_{21}} ) + \frac{1}{2 e^2 l_s^5} (x_1-x_2)^2 (|Y_{21}|^2 + |Y_{12}|^2), 
%\end{split}
%\end{align}
We have separated the Hamiltonian into ``slow" and ``fast" parts.  The interaction terms make the off-diagonal elements of $Y$ massive, with a mass that scales as the separation between the two branes. For the states localized around sufficiently well-separated values of $x_i$, modeling D-branes separated by sufficiently large distances, the diagonal degrees of freedom have slow dynamics compared to that of the ``heavy" off-diagonal matrix elements, which justifies the use of the Born-Oppenheimer approximation to construct low-energy states (though as we will see, we will need to include fermionic degrees of freedom to make this approximation quantum-mechanically consistent.). We also note that the off-diagonal components $Y_{ij}$ and $Y_{ji}$ are not independent degrees of freedom as they are complex conjugates of each other. 

Note that this Hamiltonian has a large global symmetry group, corresponding to the phase rotation of each field $Y_{ij}$. Recall that these complex fields are string fields; we can take excitations with positive charge under this summetry to be ``strings" and with negative charge to be ``anti-strings".

\subsubsection{Example: $N=2$} \label{subsec:Nequals2Example}

Let us begin with the simple case of $N=2$, modeling 2 D0-branes. Here $H_{fast}$ describes a single complex oscillator:
%\textcolor{red}{HRH:The singlet constraint further restricts this to be a real harmonic oscillator}.
\be
H_{fast} = l_s e^2 \pi^{\dagger}_{Y_{12}}\pi_{Y_{12}}+ \frac{1}{e^2 l_s^5} (x_1-x_2)^2 |Y_{12}|^2.
\ee
The interaction term shows that the potential energy is independent of $y_i$, while for $x_i$, only the relative separation $x_1 - x_2$ matters. These are consistent with the expectation that the physics of D0-branes in flat space should be invariant under target space translations. Note that the $y_i$-independence of the action is just a feature of our gauge choice. We could have chosen a gauge where we diagonalize the matrix $Y$ instead, and potential would be independent of $x_i$ but dependent on $y_1 - y_2$. 
%Moreover, since the diagonal elements of the two matrices correspond to the physical positions of the D0-branes in the dual theory, we expect that the observations we make about the entanglement between the two branes in this gauge, will lift to gauge-invariant statement physical statements. 

The next step is to construct interesting states with respect to which we compute entanglement entropies for appropriate subsets of the degrees of freedom. We will consider models of low-energy states of D0-branes. 
For such fixed $x_i = x_j$, $Y_{12}$  is a complex oscillator which has charges $\pm (1,-1)$ under the action of $U(1)_1\times U(1)_2$.  When $|x_1 - x_2|$ is sufficiently large, the variables $x_i$ have slow dynamics compared to $Y_{12}$, and we may use the Born-Oppenheimer approximation.  For fixed $x_i$ the
instantaneous energy eigenstates can be labelled as $\ket{n_+,n_-; x_1 - x_2}$, where $n_{\pm}$ are the oscillator numbers with equal and opposite $U(1)\times U(1)$ charge.

After the aforementioned gauge fixing, we must still impose the residual $U(1)^2 \times S_2$ gauge invariance.  In particular, we need to write down wavefunctions that are odd under the $S_2$ action
\begin{eqnarray}
	x_1 & \leftrightarrow & x_2 \nonumber\\
	y_1 & \leftrightarrow & y_2 \nonumber\\
	Y_{12} = Y_{21}^* & \leftrightarrow & Y_{21} = Y_{12}^*
\end{eqnarray}
In this simple case, $U(1)^2$ invariance simply means that $n_+ = n_- = n$, and we will collapse these to describe the instantaneous states as $\ket{n, x_1 - x_2}$. This also guarantees that the off-diagonal sector is even under $S_2$; thus, the diagonal sector is odd. In essence the D-brane can be treated as fermions.

Thus, we can write low-energy states of the total Hamiltonian \eref{BO-Hamiltonian} as:
%Throughout, we shall assume that we are at weak coupling i.e, $ \frac{e^2}{|X_1-X_2|^3} <<1$. We begin with the following wavefunction:
\be \begin{split} \label{BO-wavefn}
	|\psi \rangle &= \frac{1}{\sqrt{2}} \int dx_1 dy_1 dx_2 dy_2  \Big( (\psi_0(x_1,x_2;y_1,y_2)  |x_1,x_2,y_1,y_2 \rangle_s |0; x_1-x_2 \rangle_f \\ &+ \sum_{n >0} \delta \psi_{n} (x_1,x_2,y_1,y_2) | x_1,x_2,y_1,y_2 \rangle_s |n;  x_1-x_2 \rangle_f) - (x_1,y_1 \leftrightarrow x_2,y_2) \Big),
\end{split}
\ee 
Again, we have absorbed both the Vandermonde determinant and the integral over the gauge group in the amplitudes  $\psi_0$,$\delta\psi_n$ (see appendix \ref{MatrixMeasure} for more details). These amplitudes can be general 2-body wavefunctions with unit norm; we write out the antisymmetrization explicitly, with the factor of $1/\sqrt{2}$ out front yielding the correct norm for the full theory. That said, for the purposes of computing target space entropy, o, we can simply fix gauge and keep only the first term in the above sum so long as $x_1 \neq x_2$, dropping the factor of $1/\sqrt{2}$

In the limit that $|x_1 - x_2| \ll l_s$, we expect $\psi_{n>0}$ to be small compared to $\psi_0$, of order $\sim \frac{e \ell_s^{3/2}}{\delta x^{3/2}}$ for eigenvalues separated by $\delta x \equiv |x_1 - x_2$ (see Appendix \ref{sec:BOCorrections}). At leading order, we set $c_n = 0$. To find the dynamics of the slow modes, we first solve the Schr\"{o}dinger equation for the fast modes with fixed values of $x_i, y_i$ and compute the energy $E_0(x_1 - x_2)$ of the instantaneous ground state; this appears as an additional potential term in the effective Hamiltonian for $\psi_0$,
At this point we have a problem.  The linear potential that appears in our purely bosonic two-matrix model means that there are no good well-separated states with energies low compared to the oscillator modes. This problem can be cured by supersymmetry; the fermionic degrees of freedom contribute terms to the effective potential which cancel the static potential, leaving a weak velocity-dependent force. We will assume that such a cancellation exists, without including the degrees of freedom that lead to it.
In a weakly coupled system, we expect fermions to contribute additively to entanglement, and that the qualitative lessons regarding the role of ``fast" modes in entanglement calculations will carry over to a more precise treatment. 
%\footnote{One issue is that, unlike the BFSS model which has a non-zero Witten index, the corresponding models with $\mathcal{N} =4,8$ supercharges have vanishing Witten index. For such m and we cannot make a statement about the existence of zero-energy ground states \cite{Yi_1997} for a supersymmetric completion of our 2-matrix model.} 

Now that we have set up the Born-Oppenheimer framework, we will describe our prescription to compute entanglement entropy for branes localized at $Y_{(1)} < 0$. We project our space onto a subspace of the Hilbert space for which one of $x_i$ is negative and the other positive. Once this is done, the upshot of the discussion in our previous section is that we can fix the permutation gauge and do our calculation with one image of the permutation group $S_2$ for which $x_1 < 0$. We next need to make a decision as to whether or not the degrees of freedom $Y_{12}$ are to be included in the observable sector. Following the discussion in \S3.1, we assume they are not visible.
%-- said another way, we assume that the density matrix $\R_1$ is built entirely from the Heisenberg algebra generated by $x_1, \pi_{x_1}, y_1, \pi_{y_1}$ up to actions of the permutation group on the full state, that is $\R_1 = \R_1 (x_1,\tilde{x}_1,y_1, \tilde{y}_1)$. In this class of states we will refer to the variables $x_i,y_i$ with $x_i < 0$ as the coordinates as the ``first" brane and the other coordinates as the ``second" brane.

%Our prescription for computing entanglement between two branes is an example of target space entanglement (see \cite{MazencRenard:2019} for a recent discussion of target-space entanglement in the $c=1$ matrix model). 

%In the most general computation of entanglement entropy in this setup, $\R_{1}$ breaks up into superselection sectors labeled by particle numbers. However, we will closely follow our discussion of identical-particle entanglement in section \ref{sec:Identical particle entanglement} and project the fully antisymmetrized wavefunction of two branes onto the superselection sector where only one of the branes lives in region $1$ i.e, we ignore the superselection sectors where both the branes or no branes live in region $1$. Henceforth, we will refer to the D0-brane living in `region 1' as the `first' brane and the one living in `region 2' as the `second' brane. 

%Even though we do not have to deal with particle-number superselection sectors, 
For the case of general $N$, after we project on a fixed number of $X_{(1)}$ eigenvalues satisfying $x_i < 0$, the density matrix breaks up into superselection sectors corresponding to the $U(1)^n$ flux generated by excitations $Y_{ij}$ with $x_i < 0$, $x_j > 0$. We will discuss this further below. For the case of $N = 2$ with one negative eigenvalue, $U(1)^2$ invariance means this flux is always vanishing. In this sense the $N=2$ example misses an important part of the entanglement calculation for large-$N$ theories. Nonetheless, it isolates and captures another important feature, entanglement induced by $Y_{12}$, so we will proceed.  

%the computation of $\R_1$ leads to a different set of superselection sectors labeled by the total visible $U(1)$ charge entering each side. These are nothing but the number of excitations of the fast modes of the string $Y_{12}$ i.e, the value of $n$ in \eref{BO-wavefn}. We will first explain these superselection sectors in our example of two branes in the two-matrix model and later comment on its presence in any matrix model.

Let us now try to study the entanglement between the ``first" and the ``second" brane when the fast modes are placed in the instantaneous ground state $\ket{0,x_1 - x_2}$, so that we are (assuming the existence of additional fermionic modes which cancel the zero point energy) studying low-energy states of the system at leading order in the Born-Oppenheimer approximation. We will suppress the $Y$-coordinates throughout.  We further assume that we can write low-energy states for which 
$\psi_0(x_i,y_i) = \psi_1(x_1,y_1) \psi_2(x_2,y_2)$. Our essential point is that the $x_1 - x_2$ dependence of the instantaneous ground state of the heavy modes already induces nontrivial entanglement..

Tracing over both the center of mass coordinates $X_2,y_2$ of the second brane, and the string field $Y_{12}$, the reduced density matrix for $x_1$ is 
%\be
%\rho_1 = \int dx_1 d \tilde{x}_1 dx_2 d \tilde{x}_2 dx_2' \psi (x_1) \psi^{*}(\tilde{x}_1) |x_1 \rangle \langle \tilde{x}_1 | \psi(x_2) \psi^{*} (\tilde{x}_2) \langle x_2' | x_2 \rangle \langle \tilde{x}_2 | x_2' \rangle tr_{Y_{12}} (|0; x_1-x_2 \rangle \langle 0; \tilde{x}_1 -x_2| ).
%\ee 
\be
\rho_1 (x_1,\tilde{x}_1) =  \int dx_2 \psi_1(x_1) \psi_1^{*}(\tilde{x}_1) |\psi_2(x_2)|^2   \sqrt{\frac{2}{x_1-2 x_2+\tilde{x}_1}} (x_1-x_2)^{\frac14}(\tilde{x}_1-x_2)^{\frac14}
\ee
(We ignore the $y_1$-dependence as in our gauge, $y_1$ does not couple to $Y_{12}$ and the wavefunctions for $x_i,y_i$ are unentangled).  We can easily  check that the trace of the reduced density matrix defined above is one for appropriately normalized wavefunctions $\psi$.  An additional factor of 2 arising from the two terms in the antisymmetrized wavefunction give the same contribution to the density matrix is cancelled by a factor of $1/2 = 1/\sqrt{2}^2$ from the normalization of the appropriately symmetrized wavefunction. 

The kernel $K(x_1,\tilde{x}_1,x_2) =  \sqrt{\frac{2}{x_1-2x_2+\tilde{x}_1}} (x_1-x_2)^{\frac14} (\tilde{x}_1-x_2)^{\frac14}$,
arises from the $x_1 - x_2$ dependence of the state $\ket{0,x_1 - x_2}$ of the ``heavy" modes that we have declared unobservable.
In the present case this yields non-trivial entanglement even for the case that $\psi_0$ factorizes. In order to see this let us compute the second Renyi entropy, also known as the purity of the state. This is simpler to compute than the von Neumann entropy but is a good a probe of the existence of entanglement. We choose for simplicity $\psi_1(x_1) = \frac{1}{\sqrt{2}}\Big(\frac{1}{2\pi \sigma^2 }\Big)^{\frac14} e^{-(x_1+d)^2/4\sigma^2} , \quad |\psi_2(x_2)|^2 = \frac12 \delta (x_2-d)$. Technically, note that $\psi_1$ has support on both sides of the entangling surface; however, this part of the wavefunction is 
exponentially small and we will ignore its effects. Also, technically the delta function for $x_2$ carries infinite energy; we are simply using this as a model for a well-localized wavefunction to simplify our analysis. The essential point is that the wavefunction for the diagonal components has support over a range of values of $x_1 - x_2$. Since the state of $Y_{12}$ depends on this separation, $Y_{12}$ becomes entangled with $x_1$. 

In the end, in our approximation, the purity becomes:
\be
tr(\rho_1^2) =  \int  \frac{dx_1 d \tilde{x}_1}{2 \pi \s^2} exp{\Bigg(-\frac{(x_1+d)^2+(\tilde{x}_1-d)^2}{2\s^2} \Bigg)} K^2 (x_1,\tilde{x_1},d),
\ee
% 
%where we have centered the first brane around $x_1= -d$ and placed the second brane at $x_2=d$. 
The integral can be evaluated using the saddle-point approximation if $\sigma \ll d$. Performing the $\tilde{x}_1$ integral first, 
\begin{eqnarray}
tr(\rho^2) & = & 2 \int_{-\infty}^{0}  \frac{dx_1}{\sqrt{2\pi \s^2}}  e^{-\frac{(x_1+d)^2}{2\s^2} } (x_1-d)^{\frac12}  \Bigg[g(x_1,-d)+ \frac{\s^2}{2}  \frac{\partial^2 g}{\partial{\tilde x}_1^2}(x_1,-d)+ O \Bigg(\frac{\s^4}{d^4}\Bigg) \Bigg] \ ;\nonumber\\
& & \qquad \qquad g(x_1,{\tilde x}_1) = \frac{(\tilde{x}_1+d)^{\frac12}} {\frac{x_1+\tilde{x_1}} {2} -d}. 
\end{eqnarray}
We have an overall factor of two as we are integrating over half the real line. Upon performing the integral over $x_1$, again in the saddle point approximation, we find
\be \label{purity_twobranes}
tr(\rho^2)= 1-\frac{\s^2}{32d^2} + O\Bigg(\frac{\s^4}{d^4} \Bigg).
\ee  
We have retained only the leading saddle $x_1\sim -d$ in this integral. Note that this expression for the purity is something we could have guessed by dimensional analysis except for the numerical factor in the second term. 

We see that the the state is mixed as long as one of the branes is not held fixed at a location (if both branes are held fixed, $\s \rightarrow 0$ and there is nothing to decohere). This entanglement %we see at leading order in the Born-Oppenheimer, even when the strings are in their instantaneous ground state, 
is induced by the off-diagonal degrees of freedom $Y_{12}$. From the point of view of the D-branes,
entanglement between clumps of D-branes will be induced in part by the dependence of the state of the ``off-diagonal" strings, stretching between these clumps, on the location of said clumps.

As we have stated, our states are only parametrically lighter than the mass of the off-diagonal modes if we are able to somehow
cancel off the contribution of these modes to the effective potential for $x_i$. In the BFSS model, this is a consequence of supersymmetry. In these cases, the dependence of the instantaneous ground state of the off-diagonal fermionic degrees of freedom will also contribute to the kernel $K$. We expect this will not cancel the entanglement induced by $K$. In simple free field examples,
fermions contribute additively to entanglement, for example in the case of spatial entanglement in a 2d CFT, which scales with the
central charge of the theory \cite{Holzhey:1994we,Calabrese:2004eu,Calabrese:2009qy}.

A more accurate calculation would include corrections at sub-leading order in the Born-Oppenheimer approximation: that is, corrections to the adiabatic approximation to the state of the off-diagonal modes. As explained in the appendix \ref{sec:BOCorrections}, these corrections lead to even, pairwise excitations of the strings which leads to zero net flux on each string; a simple consequence of imposing invariance under the 
$U(1)^2$ gauge group which remains after we diagonalize $X_{(1)}$. 

Qualitatively, new effects occur when we extend our analysis to the $U(N > 2)$ ,with $k < N$ visible branes in the region $Y^{(1)} < 0$. In this case, the density matrix will generally break up into ``superselection sectors" as we have discussed above, corresponding to components of the quantum state that contain different configurations of unobserved off-diagonal matrix elements which carry nontrivial $U(1)^k$ charge. To understand these better, 
we will pass in the next section mapping our target space entanglement calculations onto a computation of entanglement in a gauge theory living on a complete lattice. Once we have explained this map, we return to the discussion of three matrices in BFSS theory which we will argue captures all the qualitatively essential features of target space entanglement in the full BFSS theory.

%Thus, these corrections do not change the `superselection' sector of the ground state and hence, do not change the entanglement structure of the ground state from \eref{purity_twobranes} to leading order in $\frac{\s^2}{d^2}$. 

. 

%%%%%%%%%%%%%%%%%%%%%%%%%%%%%%%%%%%%%%%%%%%%

%%%%%%%%%%%%%%%%%%%%%%%%%%%%%%%%%%%%%%%%%%%%
\subsection{$N > 2$: superselection sectors} \label{EHS}

For $N>2$, we can have excitations of off-diagonal matrix elements such that the state of a given matrix element carries non-vanishing charge under a $U(1)\times U(1)$ subgroup of $U(1)^N$, while the total $U(1)^N$ charge is vanishing. In particular, this means that the density matrix for $k$ `visible' eigenvalues breaks up into terms labeled by the total $U(1)^k$ charge carried by excitations charged under $U(1)^k \times U(1)^{N-k}$. 

Let us explore this phenomenon further. Before projecting onto the $S_N$-symmetric states -- or if the eigenvalues of $X_1$ are distinct, after fixing the permutation symmetry by ordering the eigenvalues -- the matrix quantum mechanical model can be treated as a $U(1)$ lattice gauge theory on a `complete' (all-to-all) N-site lattice, in the Extended Hilbert Space formalism of \cite{TrvediSoni:2015}. A natural physical treatment of this system, motivated by theories of D0-branes, leads to an `electric center' prescription for treating the degrees of freedom that connect the `hidden' and `visible' degrees of freedom. 
%
%In this section we map the computation of target space entanglement in D0-brane quantum mechanics discussed in the previous section \ref{TwoMatrixEntanglemnent} to the computation of entanglement in Hamiltonian lattice gauge theories in the Extended Hilbert Space formalism (EHS) of \cite{TrvediSoni:2015}.  As we have explained before, this discussion is motivated by the observation that the instantaneous states $ |n; x_1-x_2 \rangle $ of the diagonal degrees of freedom transform as bi-fundamental of $U(1) \times U(1)$. Throughout the discussion, we will work with the `electric centre' definition in the EHS formalism. 
We begin with $N$ lattice points and $N^2$ links between them. The diagonal matrix elements $X_{(i)kk}$ `live' on the $k$th lattice site, while the off-diagonal terms $X_{(i)kl}$ live on the links joining the $k$th and $l$th lattice sites. Each site $k$ has an unbroken gauge symmetry $U(1)_k$ associated to it, with the diagonal matrix elements being neutral and the complex fields $X_{(i)kl}$ having charge $\pm(1,-1)$ under $U(1)_k\times U(1)_l$. Each link can carry nontrivial $U(1)^2$, so long as adjacent links also carry charge in such a way that the total $U(1)^N$ charge is zero. We can clearly organize the Hilbert space according to the amount of charge flowing through each link, with `charged' links joining to form closed loops.  Note that for the example discussed in the previous section, we have a lattice with two vertices representing the coordinates of the two branes; the total $U(1)$ charge associated to the edge must have $n_{+}-n_{-}=0$ (in string theory language, excitations will come in string-anti string pairs).

The upshot is that the Hilbert space is the $U(1)^N \rtimes S_N$-invariant subspace of the ``extended" Hilbert space
\be
\mathcal{H} = \otimes \mathcal{H}_{V_{i}} \otimes \mathcal{H}_{ij}\ ,\label{eq:extended}
\ee
where $\mathcal{H}_{V_i}$ is the local Hilbert space at each vertex and $\mathcal{H}_{ij}$ is the local Hilbert space on each link. Invariance under $U(1)^N$ transformations leads to the conservation of total $U(1)$ charge at each vertex. We must further impose the $S_N$ permutation symmetry of our matrix theory. We can impose this by writing states in a basis of $S_N$ orbits of $U(1)^N$-invariant wavefunctions on $\mathcal{H}$. This extends the ``unentangled" basis in \S2. Alternatively, if the eigenvalues of $X_1$ are distinct, we can choose a particular element of this orbit, for example the obe such that $X_{11} < X_{22} < \ldots X_{NN}$. Target space entanglement is gauge-invariant and we can use either treatment of $S_N$ to calculate it.

%which will relabel both the edges and the vertices. However, we know that the contribution of each term in the fully antisymmetrized wavefunction of the matrix theory to the density matrix is the same (see \ref{TwoMatrixEntanglemnent}). Therefore, we will instead work with a given representative of the equivalence class generated by the the permutation symmetry acting on the complete lattice.

Let us now consider $k$ of the $N-k$ lattice sites, and the associated links between them, to be `visible', modeling for example D-branes that lie on one side of an entangling surface, and the string fields for open strings stretching between them. We can now trace over the degrees of freedom on the $N-k$ `hidden' vertices, as well as the corresponding links, as well as the string fields living on the links connecting the hidden and visible vertices. Since this a complete lattice all vertices are also boundary vertices. The resulting Hilbert space and the corresponding reduced density matrix break up into superselection sectors
\be
\mathcal{H}_{in} = \bigoplus \mathcal{H}^{\vec{q}}_{in} \quad \rho_{in} = \bigoplus \rho^{\vec{q}}_{in},
\ee 
where the reduced density matrix $\rho_{in}$ is block diagonal and $\vec{q} = q_1, \cdots q_k$ is the vector that denotes the total $U(1)^k$ charge due to excitations on links joining the visible and hidden vertices. The entanglement entropy associated to this state is given by
\begin{equation}
S_{EE} = -\sum_{\vec{q}}p_{\vec{q}}\log{p_{\vec{q}}}-\sum_{\vec{q}}p_{\vec{q}}\tr[\rho_{in}^{\vec{q}}\log{\rho_{in}^{\vec{q}}}].
\end{equation}
The first term, which we refer to as the classical piece, is given by the entropy due to the probability to be in a particular superselection sector. The second piece, which we refer to as the quantum piece, is given by a weighted sum of the entanglement entropy of each superselection sector. No gauge invariant operator can act to change the superselection sector unless it acts on degrees of freedom both inside and outside the spatial cut (namely a Wilson loop crossing the boundary); in particular no observable supported on the ``visible" degrees of freedom can
change ${\vec q}$.

%Lastly, we also note that fermions can be incorporated in this picture by considering  extra links that correspond to `anti-strings' stretching between the branes. 

We now proceed to illustrate our map in a simple example.

\subsubsection{$N = 4$ example}

Let us return to the case of two matrices. We might ask whether corrections to the leading-order Born-Oppenheimer approximation would induce states with string excitations forming loops on the lattice, such that the net $U(1)^2$ gauge charge on each link is nonzero. However, as noted previously, the Hamiltonian \ref{BO-Hamiltonian} has an enhanced global symmetry corresponding to $Y_{ij} \to e^{i\phi_{ij}} Y_{ij}$. A single "string" associated to a given link carries nonzero charge under this symmetry. If the leading term in the Born-Oppenheimer approximation corresponds to the instantaneous vacuum, which carries vanishing charge, corrections to adiabaticity from $x_i$ being diagonal will not break this symmetry, and the aforementioned loop configurations will not be induced. 
In the language of D0-branes, open string would be created in string-antistring pairs, so that the excitations on each link carry vanishing gauge charge.

As we will see below, for three or more matrices, the Hamiltonian no longer has this symmetry and even the instantaneous ground state for the heavy off-diagonal modes will contain components with loops of strings.

However, for illusrative purposes, we simply construct a state in the two-matrix thwory, for which the reduced density matrix for visible branes breaks up into superselection sectors labeled by the charge carried by the hidden degrees of freedom. This will be some highly excited state in the 2-matrix theory.

% We might ask whether higher orders in the Born-Oppenheimer approximation can induce non-trivial string excitations forming loops around the lattice, with nontrivial $U(1)^2 \in U(1)^N$ charges carried by individual links. This does not happen at sub-leading order. 

% We find that simply do not have the necessary operators to create these loops at leading and sub-leading order. In what follows, the specific charge configuration considered is an example of what can arise at higher orders in the Born-Oppenheimer approximation.

\begin{figure}[H]
	\centering
	\includegraphics[width=.6\textwidth]{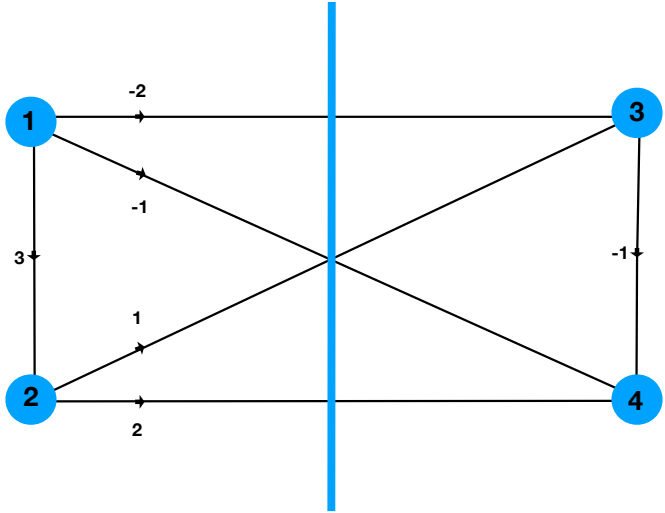}
	\caption{\label{fig:latticeD0} A string configuration for the bipartition of 4 D0-branes.
	}
\end{figure}

Let us consider 4 ``D0-branes" which are split into two groups separated by a large distance $d$, as illustrated in Figure (\ref{fig:latticeD0}). In order to focus on the classical contribution we will consider the situation where we have pinned down the branes to unique locations ${\vec Y}_1,\ldots,{\vec Y}_4$, where ${\vec Y}_k = (X_{kk},Y_{kk})$ is a two-dimensional vector of diagonal matrix elements which we can think of as D0-brane position, and
we are working in the gauge with diagonal $X$ as before.
We take ${\vec Y}_1,{\vec Y}_2$ to lie on on the visible side of the entangling surface: that is, $X_{11}, X_{22} < 0$. In the figure, the sites of the lattice correspond to the D-branes; we have denoted the associated location by the number $1,\ldots,4$ corresponding to the subscript of ${\vec Y}_I$, The links denote the ``off-diagonal" matrix elements, labeled by position subscript at each end. The arrows between sites $i$ and $j$ define the meaning of the sign of the charge $\alpha_{ij}$; an excitation with charge $\alpha_{ij} = +1$ on a given link carries positive charge at site $j$ at the head of the arrow, and negative charge at site $i$ at the tail of the arrow. The Figure denotes a particular configuration of off-diagonal excitations which satisfies $U(1)^4$ gauge invariance; that is, the total charge at each vertex is zero.

%Let us consider 4 D0 branes which are split into two groups separated by a large distance $d$ and represent the resultant situation on a complete lattice. Since classical entanglement is nontrivial to generate we will consider the situation where we have pinned down the branes to unique locations which can then be assigned position labels. In this picture as explained above we can think or the branes as being on the vertices of a complete lattice of a $U(1)$ gauge theory. We can then pick a simple excited state of the theory which represents a configuration of strings with each string in a single excited state. While one could consider more complicated states with inherent quantum entanglement, like states where each link on the lattice corresponds to a superposition of many excitations of the corresponding string, we will not consider them here. 

If the vertex labels correspond to target space position, the graph in Figure (\ref{fig:latticeD0}) then corresponds to a specific $U(1)^4\times S_4$-invariant state. As before we can describe this in a manifestly $S_N$-invariant fashion, or by fixing the $S_N$ gauge symmetry. In the former case, we start with the state 
\be\label{eq:seedstate}
\ket{\psi} = \ket{{\vec Y}_1,\ldots,{\vec Y}_4} 
\ket{\alpha_{12} = 3,\alpha_{13} = -2,\alpha_{14} = -1,
\alpha_{23} = 1,\alpha_{24} = 2,\alpha_{34} = -1}
\ee
in the ``extended" Hilbert space \ref{eq:extended} and then sum over the $S_N$ orbit that includes this state, with appropriate signs:
\be
|\psi \rangle := \frac{1}{4!}\sum_{\sigma \in S_4} sgn(\sigma)\ket{{\vec Y}_{\sigma(1)}\ldots{\vec Y}_{\sigma(4)}} \ket{\{\alpha_{\sigma(i)\sigma(j)} \}}
\label{eq:unent}
\ee
For example, the contributions to the sum from $\sigma = {\bf 1}$ and $\sigma = (12)$ are:
\be
|3,-2,-1,1,2,-1\rangle|1,2,3,4\rangle  -|3,1,2,-2,-1,-1\rangle|2,1,3,4\rangle.
\ee
Alternatively, we can fix the $S_N$ symmetry, for example by imposing the ordering $X_{11} < X_{22} < X_{33} < X_{44}$.
With this ordering we can just work directing with (\ref{eq:seedstate}) and compute entanglement in the standard fashion.

States of the form (\ref{eq:unent}) have vanishing target space entanglement. On the other hand, if we begin with the extended Hilbert
space state
\begin{equation}
   |\psi\rangle=\frac{1}{\sqrt{2}}( |3,-2,-1,1,2,-1\rangle+|4,-3,-1,2,2,-1\rangle)|{\vec Y}_1,{\vec Y}_2,{\vec Y}_3,{\vec Y}_4\rangle
   \label{eq:entkernel}
\end{equation}
and sum over the orbit of $S_4$, which acts on each term separately, or equivalently impose an ordering on $X_{kk}$, we have a state which has target space entanglement.
We perform the partial trace over all degrees of freedom except ${\vec Y}_{1,2}$, $\alpha_{12}$ to find the reduced density matrix:
\begin{equation}
    \rho_{12}=\frac{1}{4}\left(\ket{\alpha_{12} = 3}\bra{\alpha_{12} = 3} + \ket{\alpha_{12} = 4}\bra{\alpha_{12} = 4}\right)
    \ket{{\vec Y}_1,{\vec Y}_2}\bra{{\vec Y}_1,{\vec Y}_2} - (1 \leftrightarrow 2)
\end{equation}
where the $(1 \leftrightarrow 2)$ appears if we have not fixed the $S_4$ symmetry. Each term in this sum corresponds to a distinct superselection sector with different amounts of charge flowing into the visible vertices.
Each superselection sector appears with probability $\half$; the total classical entropy is thus $S = \ln 2$.

\subsection{Entanglement with three matrices} \label{sec:ThreeMartrices}

In the previous sections, we discussed the two-matrix model at length and introduced our framework for computing entanglement between two clumps of well-separated branes. Starting from the Born-Oppenheimer approximation for low-energy modes, with the off-diagonal terms in the instantaneous ground state, we found (see \ref{sec:BOCorrections} for more details) that non-adiabatic corrections to the Born-Oppenheimer approximation included only string-anti string pairs associated to each off diagonal mode (a link of the complete lattice, in the language of \S3.2.2.)

Next, we consider the case of three bosonic matrices, 
%\footnote{We should note that there is a supersymmetric matrix model called the `mini-BFSS model' with four supercharges \cite{Denef_2002,Claudson:1984th,Asplund_2016} which has vanishing Witten index \cite{Yi_1997}. Nevertheless, there is some evidence to believe that this theory has supersymmetric ground states \cite{Anous_2019}. We will continue to focus only on the bosonic version in this paper.} 
for which new qualitative features arise in the entanglement structure of low-energy states.\footnote{See \cite{Han:2019wue} for a discussion of entanglement in the low-energy states of a related three-matrix model.}
Specifically, corrections due to interactions induce components of the quantum state with excitations that carry $U(1)$ charge from the hidden to the visible degrees of freedom, inducing ``classical" terms in the entanglement entropy. We will sketch the manner in which such components appear.

We start with the following action,
\be \label{BFSSThreeMatrixaction}
S = \int dt  \frac{1}{2e^2l_s} tr \Bigg [ \dot{X}^2+ \dot{Y}^2+\dot{Z}^2 - \frac{1}{l_s^4}([X,Y]^2 - [Z,X]^2- [Y,Z]^2)\Bigg] .
\ee
As before, we can use the $U(N)$ symmetry to diagonalize the matrix $X$. Passing to the Hamiltonian, and assuming the eigenvalues of $X$ are well-separated, we separate the dynamics of the ``slow" diagonal elements and the ```fast", heavy off-diagonal elements:
\begin{align} \label{ThreeMatrixHamiltonian}
\begin{split}
H            &= H_{slow} + H_{fast}, \\
H_{slow} & = \frac{e^2 l_s}{2} \sum_{i=0}^{N}  (\pi_{x_i}^2 +\pi_{y_i}^2 + \pi_{z_i}^2), \\
H_{fast} &  = \sum_{i=0, i<j}^{N} \Bigg( e^2 l_s (\pi^{\dagger}_{Y_{ij}}\pi_{Y_{ij}}+ \pi^{\dagger}_{Z_{ij}} \pi_{Z_{ij}} ) + \frac{1}{ e^2 l_s^5} (x_i-x_j)^2 (|Y_{ij}|^2 + |Z_{ij}|^2) \Bigg) \\
& \qquad \qquad + \frac{1} {2e^2 l_s^5} [Y,Z]^2 , 
\end{split}
\end{align}
The interaction terms of $H_{fast}$ within the round braces induces masses of order ${\cal O}(|x_i - x_j|)$ for the off-diagonal terms in the $Y$ and $Z$ matrices. The important new feature is the commutator term ${\rm tr} [Y,Z]^2$. In the limit of large $|x_i - x_j|$, it can be treated perturbatively. A simple analysis of how the perturbative corrections scale -- see Appendices \ref{sec:BOCorrections}, \ref{sec:pertcorr} -- shows that this term will induce corrections to the instantaneous energy eigenstates of order $\sqrt{g} \equiv \frac{e l_s^{3/2}}{\delta x^{3/2}}$ and of order $g^2$. These small parameters also control non-adiabatic corrections to the Born-Oppenheimer approximation, although the order $\sqrt{g}$ terms are further suppressed by a ratio of energy scales.  A complete treatment will have to take care of both effects simultaneously. 
%
% However, we are interested in the case where the separation $x_i-x_j$ is large compared to any of the matrix elements in the $Y$ and $Z$ matrix. In this limit, the new interaction term can be treated as a perturbation to the original Hamiltonian:
% \be \label{ThreematrixH}
% H_{fast} = H^0_{fast} + \frac{\la} {2e^2 l_s^5} [Y,Z]^2,
% \ee
% where we have introduced a dimensionless parameter $\la$ for convenience. We consider the wavefunction in the ground state of the `fast' modes and it receives corrections both in the Born-oppenheimer expansion and in perturbation theory:
% \be \label{threematrix_wavefn}
% \Psi = \Psi_{BO_0} +( \la' \Psi_{BO_0}^{(1)}+ \la'^2 \Psi_{BO_0}^{(2)} + \cdots ) + ( \la' \Psi_{BO_1}^{(1)} + \la'^2 \Psi_{BO_1}^{(2)} + \cdots ) + \cdots ,
% \ee 
% where the superscripts refer to the corrections coming from perturbation theory. A little dimensional analysis tells us that the correction to the wavefunction coming from perturbation theory mix with the corrections coming from Born-Oppenheimer analysis i.e, $\la' = \la \frac{e^2l_s^3}{(\delta x)^3}$ (see appendix \ref{sec:BOCorrections}).  
%
%Let us now focus on the qualitative features of computing entanglement/ R\'enyi entropies in the state \eref{threematrix_wavefn} -  

%$H^0_{fast}$ \eref{ThreematrixH} acting on $\Psi_{BO_0}$ creates two kinds of strings- the $Y$-strings and the $Z$-strings. 

\begin{figure}[H]
\centering
\begin{tabular}{cc}
\includegraphics[width=.5\textwidth, valign=T]{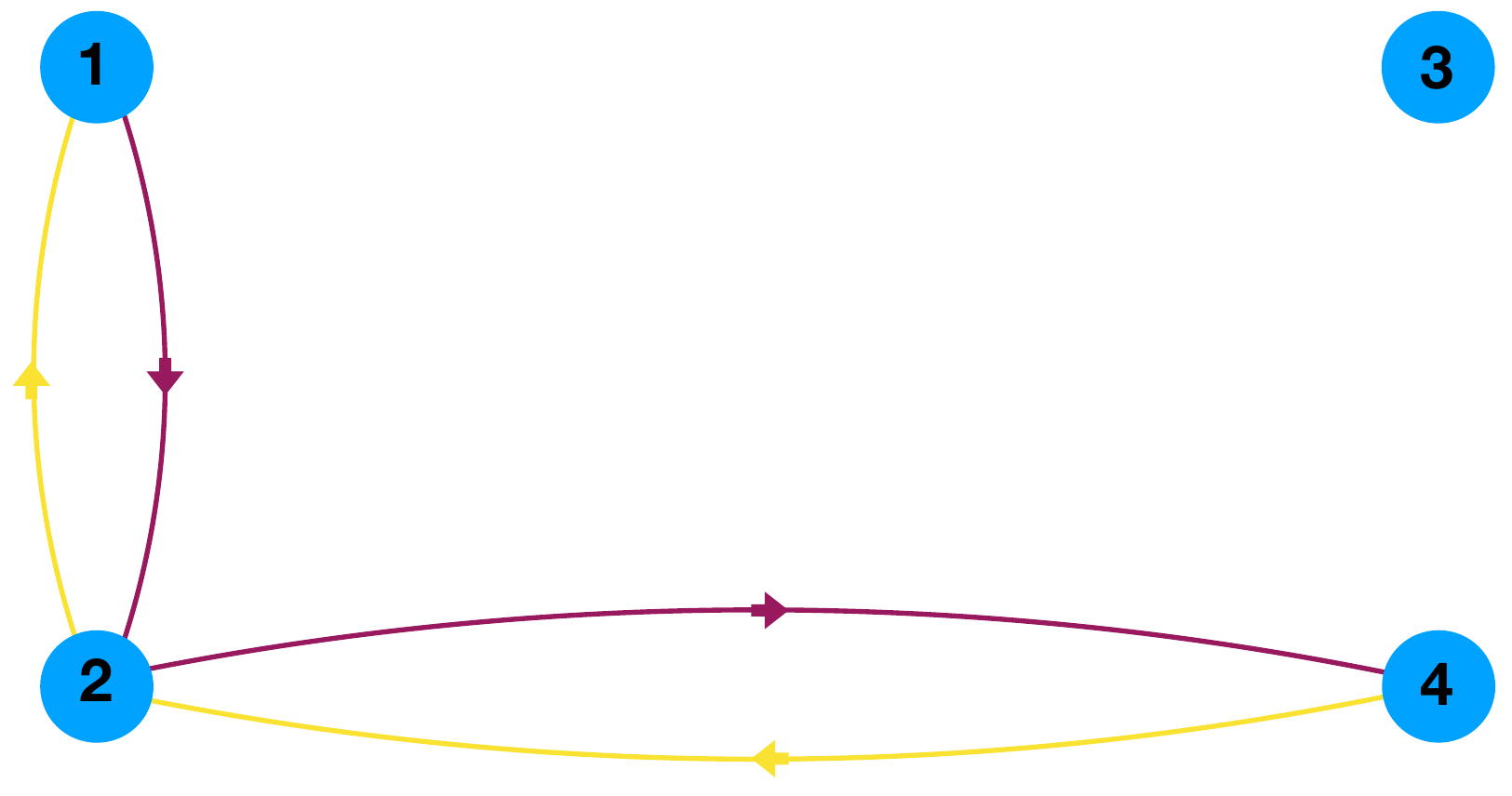} & \includegraphics[width=.5\textwidth, valign=T]{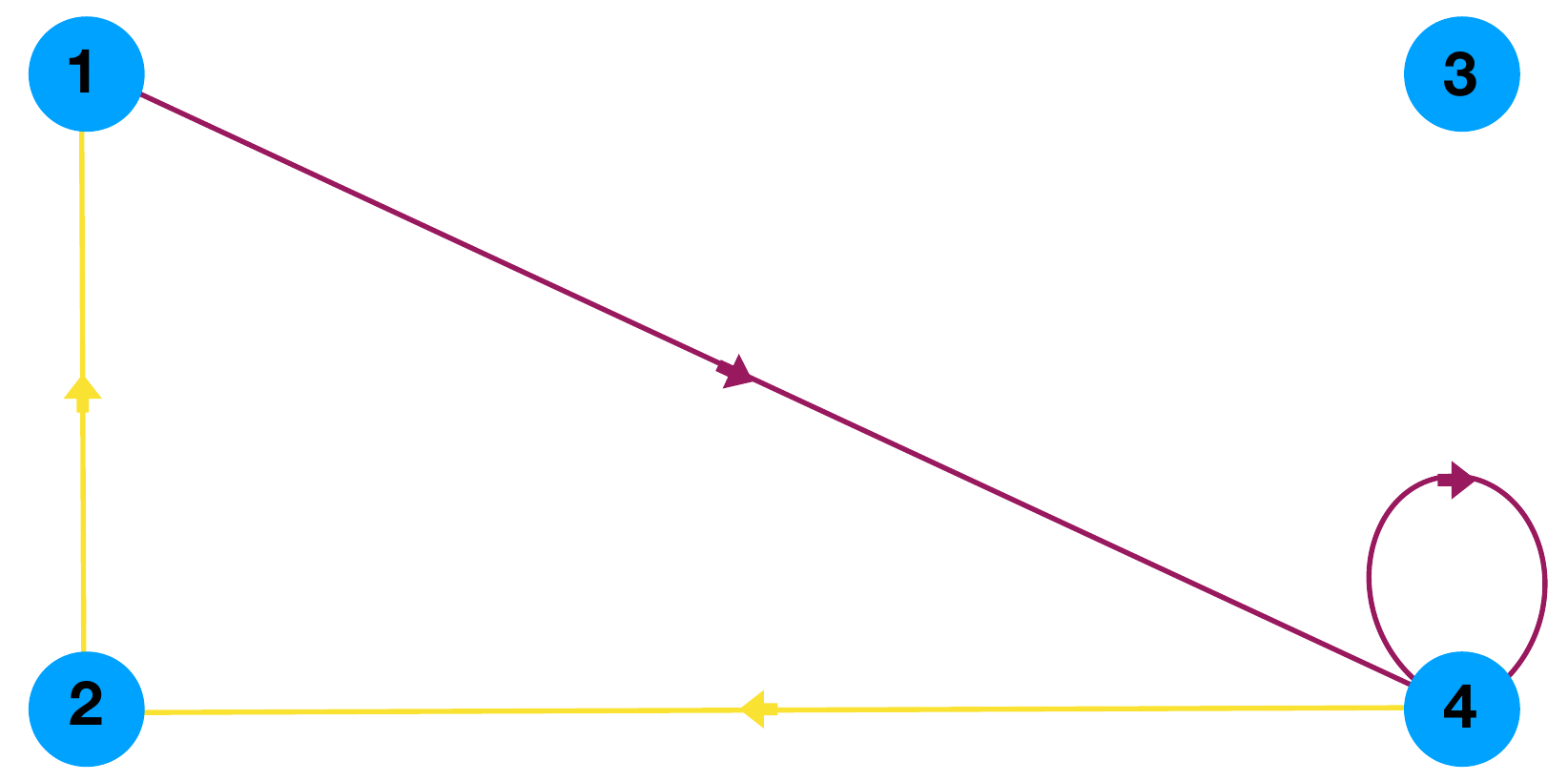} \\
$y_{12}^{*}z_{21}^{*}y_{24}z_{42}$ & $y_{12}^{*}z_{44}^{*}y_{24}z_{41}$ \\[6pt]
\multicolumn{2}{c}{\includegraphics[width=.5\textwidth]{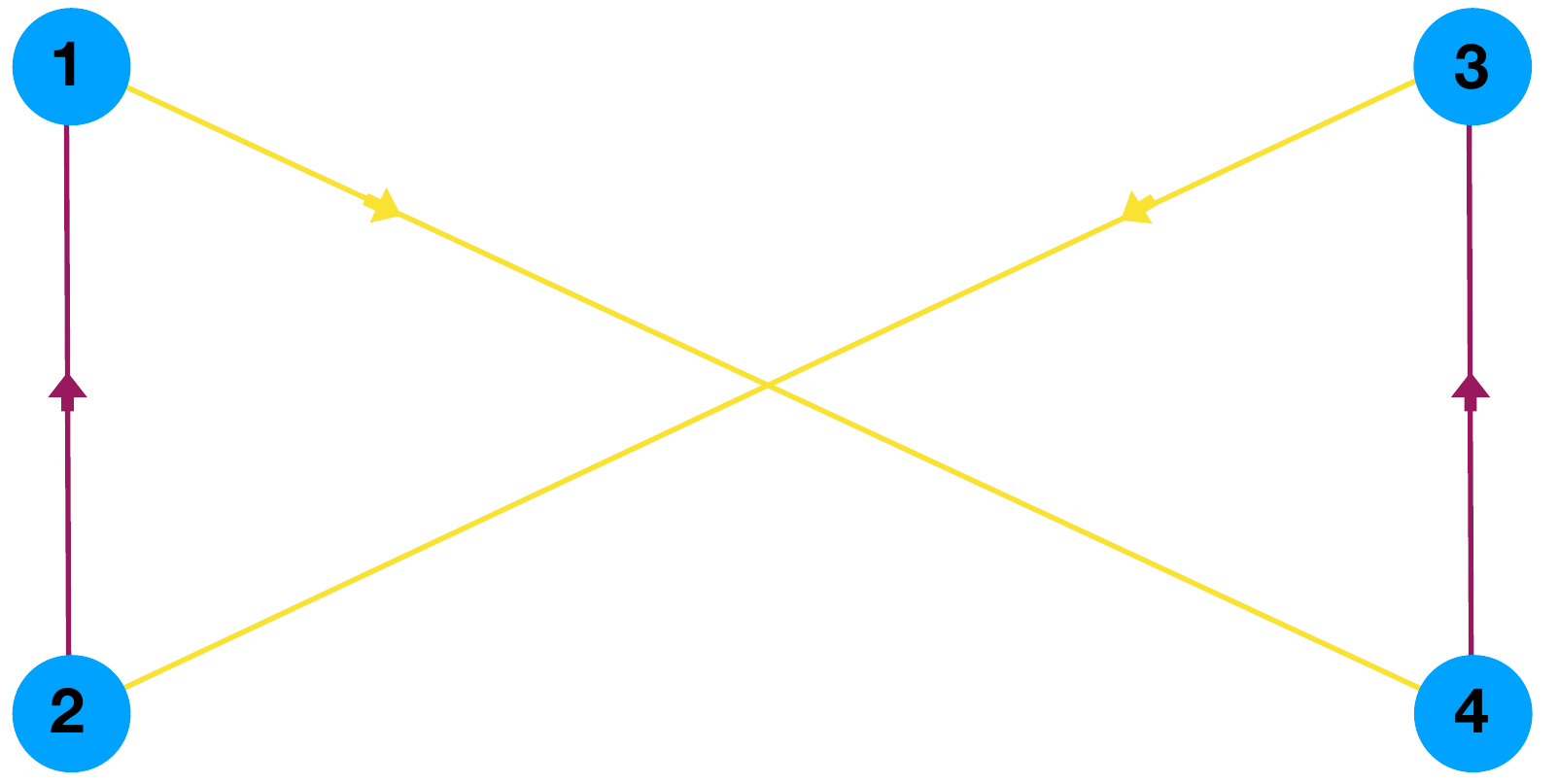} }\\
\multicolumn{2}{c}{$y_{32}^{*}z_{43}^{*}y_{41}z_{12}$}
\end{tabular}
\caption{\label{fig:3matrixstrings} Open string configurations for typical plaquette producing terms in equation \ref{BFSSThreeMatrixaction}. The Y string are shown in yellow and the Z strings in red. }
\end{figure}

The new commutator term breaks the global symmetry of separate phase rotations of the fields on each link of our complete lattice. t Thus it can and does induce corrections to the instantaneous ground state with excitations of the off-diagonal modes of $Y$, $Z$ around 3- and 4-link loops of our complete lattice: at leading order, these have single excitations on each link of the loops. When such loops connect vertex sites on each side of the entangling surface, the correction to the instantaneous ground state leads to a term in the reduced density matrix for the ``visible" degrees of freedom in which non-trivial $U(1)$ charge flows into visible lattice sites across the entangling surface. Figure 
\ref{fig:3matrixstrings}\ gives a visual picture of such corrections in the case of $N=4$, with each numerical label on the vertex denoting a target space position.

% contains many additional terms because neither $Y$ nor $Z$ is diagonal. Some of these terms correspond to operators that create rectangular loops of strings. We will refer to such operatos as `plaquette-operators'. We illustrate this using an example of four branes. Let us label them as branes $1,2,3$ and $4$ and trace over the $3$ and $4$ degrees of freedom. In figure \ref{fig:3matrixstrings} we see there are terms in the Hamiltonian that create plaquettes. These terms lead to non-trivial entanglement even when each string is in its `ground' state. 

In a generic excited state, the computation of entanglement breaks up into superselection sectors labeled by the total $U(1)$ charge at each vertex. As explained in section \ref{EHS}, we get both a quantum piece and a non-trivial classical piece. In a computation of entanglement in the full BFSS model, we expect all the features described in this section to appear. We hope to understand these aspects at a more quantitative level in a future communication.

\subsection{N scaling} \label{sec:Nscaling}

% \begin{figure}[H]
% \centering
% \includegraphics[width=.75\textwidth]{figs/Nscale3_1.pdf}
% \caption{\label{fig:Nscale3_1} A lowest order flux loop between three branes in one clump and one of the other. There are $2N_{1}(N_{1}-1)(N_{1}-2)N_{2}+2N_{2}(N_{2}-1)(N_{2}-2)N_{1}$ such configurations.
% }
% \end{figure}

% \begin{figure}[H]
% \centering
% \includegraphics[width=.75\textwidth]{figs/Nscale2_2.pdf}
% \caption{\label{fig:Nscale2_2} A lowest order flux loop between two branes in each clump. There are $2N_{1}(N_{1}-1)N_{2}(N_{2}-2)$ such configurations.
% }
% \end{figure}

We close with some preliminary comments on the scaling of the target space entanglement entropies we have been discussing, when the eigenvalues are well separated across an entangling surface. We imagine $N_1$ eigenvalues of $X$ are negative, and $N_2 = N - N_1$ positive.

First, let us revisit entanglement induced by off-diagonal matrix elements which connect matrix eigenvalues on each side of the entangling surface. We imagine a case for which the state of these modes depends on the relative separation of the diagonal modes which give them mass, as in the $N=2$ 2-matrix example given above, and not on the ``visible" off-diagonal modes. Let us further imagine a situation for which in the full quantum state, the eigenvalues on each side of the entangling surface have wavefunctions of the form $\psi_1(x_1,\ldots x_{N_1})\psi_2(x_{N_1+1},\ldots,x_N)$, and the ${\cal O}(N_1^2)$ off-diagonal modes $Y_{ij}$, $Z_{ij}$ for $1 \leq i,j \leq N_1$ are similarly uncorrelated with those for $N_1 + 1 \leq i,j \leq N$. In this case the entanglement is carried entirely by the ${\cal O}(N_1N_2)$ off-diagonal modes with $1 \leq i \leq N_1 < j \leq N$, From the standpoint of the visible branes, these modes entangle only the ${\cal O}(N_1)$ eigenvalues on one side of the entangling surface with the ${\cal O}(N_2)$ eigenvalues on the other. For a quantum system in a pure state, with $M$ visible degrees of freedom and $P$ hidden degrees of freedom, the von Neumann entropy is bounded from above by ${\rm min}(M,P)$. In this case that minimum will be $N_1$, even if $N_1 \gg N_2$. In other words, if the strings stretching across the entangling surface are not observed, they can provide a large amount of entanglement even if there are few eigenvalues with support in the ``hidden" region.

Next, we consider the ``classical" entanglement induced at weak coupling from components of low-energy states with excitations of strings that carry charge under the $U(1)^{N_1}$ gauge group for the visible degrees of freedom as well as the $U(1)^{N_2}$ gauge group for the hidden degrees of freedom. We work at lowest order in the Born-Oppenheimer approximation for which states of the off-diagonal matrix elements are taken to be their ground states. For the case of 3 matrices, the commutator term ${\rm tr} [Y,Z]^2$ in \eref{ThreeMatrixHamiltonian} induces corrections to the ground state away from the simple harmonic oscillator vacuum. As explained in section \ref{sec:ThreeMartrices}, this term creates triangular and square loops which contribute to the first-order corrections to the wavefunction:
\be
|\psi \rangle = |\psi \rangle_0 + c_t e\left(\frac{l_s}{\delta x}\right)^{3/2} \sum_{K=1}^{N_2} |(triangle)_{IJK} \rangle
+ c_s \frac{e^2l_s^3}{(\delta x)^3} \sum_{K \neq L}^{N_2} |(square)_{IJKL} \rangle + \cdots 
\ee 
where $\delta x$ is the characteristic separation between eigenvalues, and we have used the results in Appendix \ref{sec:pertcorr} to estimate the size of the corrections.

\begin{figure}[H]
   \centering
   \subfloat[  Plaquettes ]{{\includegraphics[width=.45\textwidth]{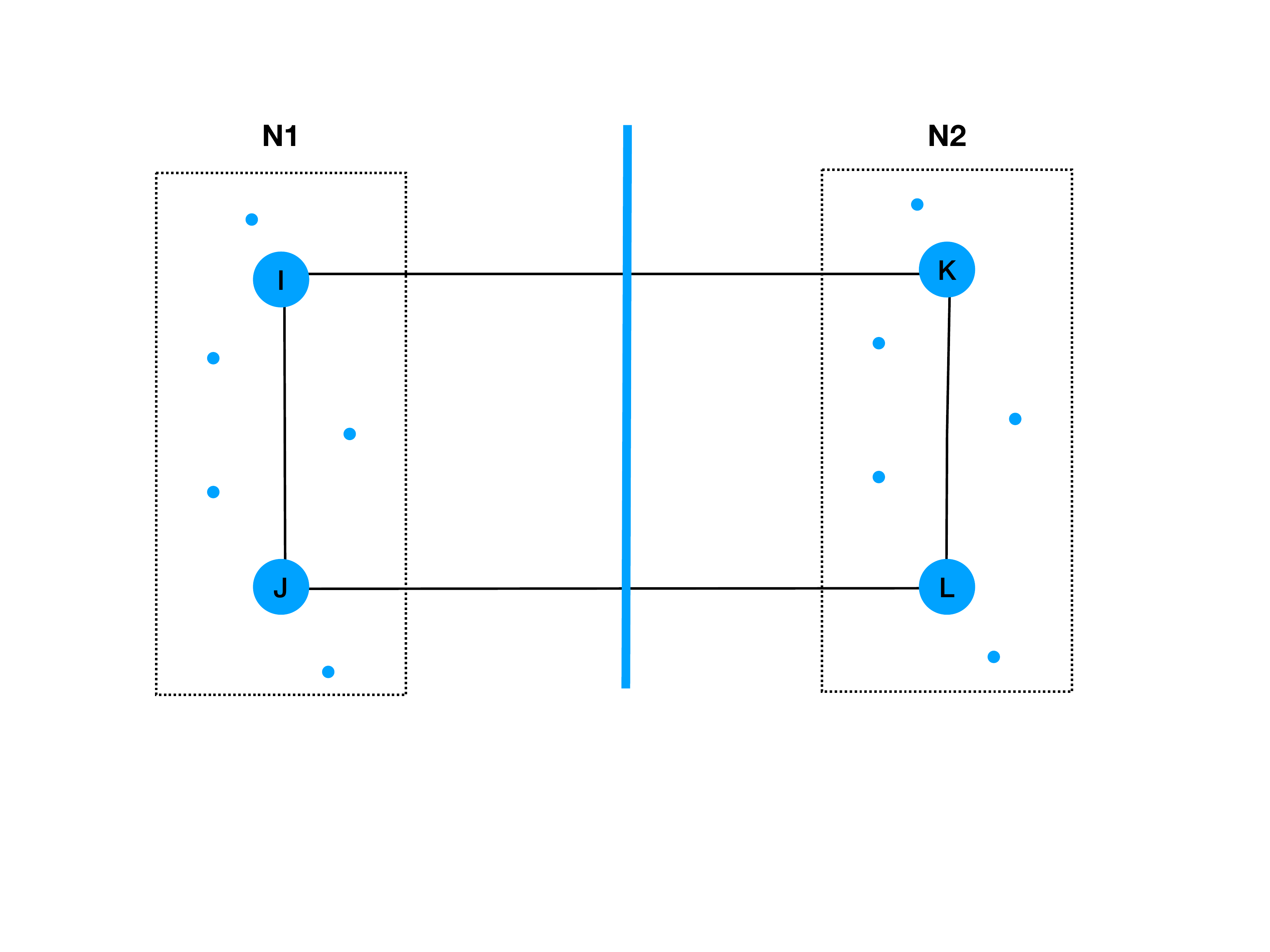}}}
    \qquad
    \subfloat[ Triangles]{{\includegraphics[width=.45\textwidth]{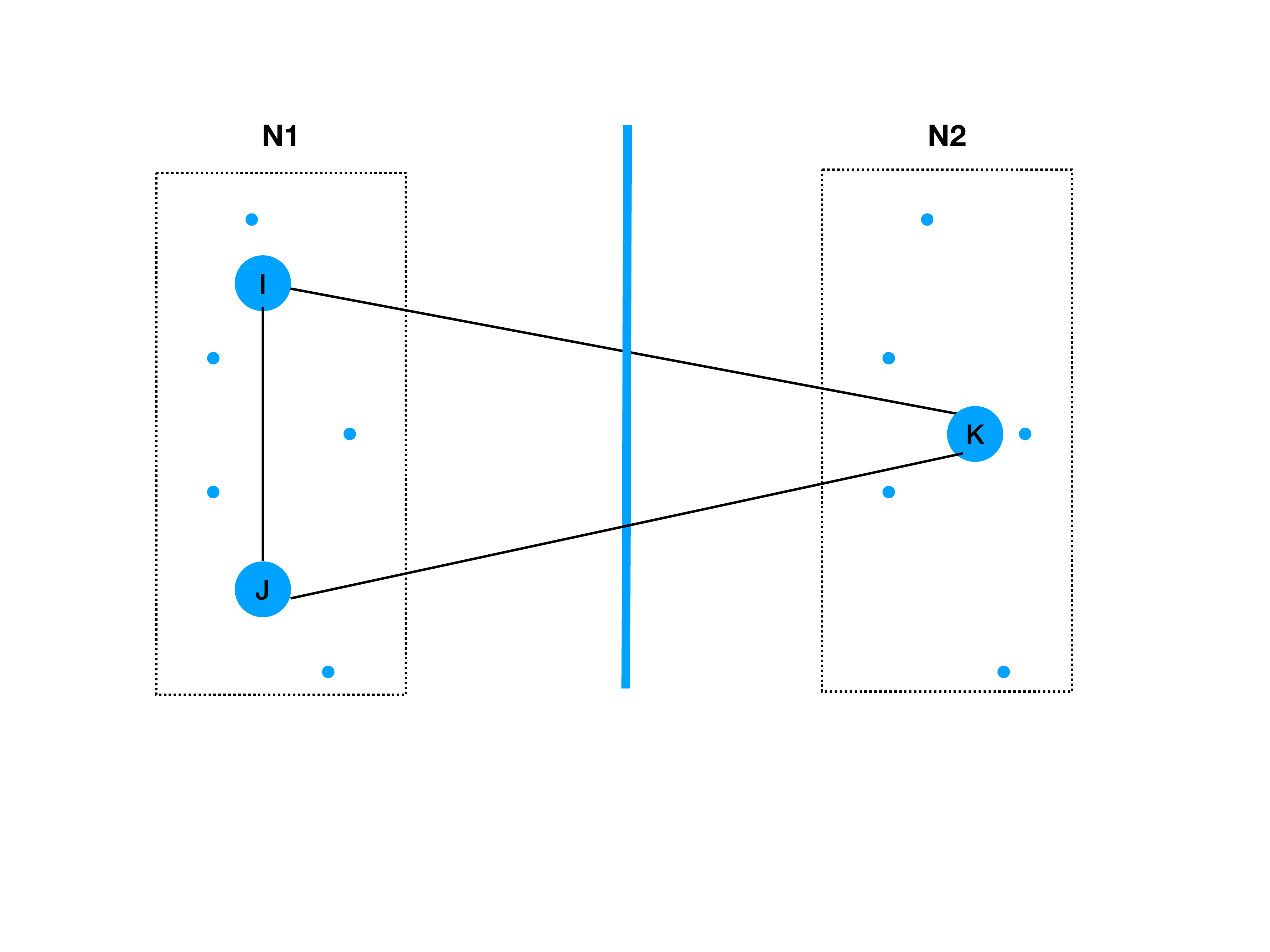} }}
    \caption{Example of a triangle and plaquette that contribute to the ground state wavefunction. The branes $I$ and $J$ are in the visible clump.}
    \label{fig:example}
\end{figure}

The pattern of excitations of off-diagonal matrix elements that appear in corrections to the ground state at leading nontrivial order are shown in Figure \ref{fig:example}. Tracing over everything except the visible degrees of freedom $(I,J)$, we obtain a density matrix of the form,
\be
\R =\R_0 + |c_t|^2 \frac{\la_2l_s^3}{\delta x^3} \R^t_{IJ}
+ |c_s|^2 \frac{\la_2^2 l_s^6}{\delta x^6}R^s_{IJ} + \cdots,
\ee 
where we have introduced the `t Hooft coupling $\la_2 = e^2N_2$. The reduced density matrix has $O(N_1^2)$ sectors, thus, in the limit $\la_2 \ll 1$, $\delta x \gg l_s$, the leading ''classical" correction the the von Neumann entropy will be of order 
\be
    S_{class} \sim - \la_2 \left(\frac{l_s}{\delta x}\right)^{3} N_1^2 \ln\left[\la_2 \left(\frac{l_s}{\delta x}\right)^{3}\right]
\ee
coming from the triangular loops.

%%%%%%%%%%%%%%%%

%  We will now discuss the case when we have two large clumps of $N_{1}$ and $N_{2}$ D0 branes. To determine the N scaling of classical term we will consider lowest energy plaquette like states with four string excitations. Such states include loops between three branes on one side of the divide (see fig. \ref{fig:Nscale3_1}) and loops between two branes on one side and two on the other (see fig. \ref{fig:Nscale2_2}). Both orientations contribute. Now to extract the scaling we can consider a maximally mixed state of all four string loop states. In this case we have classical entropy contribution
%  \begin{equation}
%  S_{\text{classical}} = \log\left(2(N_{1}(N_{1}-1)N_{2}(N_{2}-2)+N_{1}(N_{1}-1)(N_{1}-2)N_{2}+N_{2}(N_{2}-1)(N_{2}-2)N_{1}) \right)
%  \end{equation}
%  so that in the $N_{1},N_{2}$ large limit we have
%   \begin{equation}
%  S_{\text{classical}} = \log\left(N_{1}^2N_{2}^2+N_{1}^3N_{2}+N_{2}^3N_{1} \right) \approx \log(N).
%  \end{equation}
 
%  \textcolor{red}{Im not sure if this is the conclusion we want (use different state?). Do we want to adjust this to use three matrix case or leave as is?}
 
%%%%%%%%%%%%%%%%%%%%%%%%%% 
\section{Future directions}\label{sec:future}
% 1) A path-integral derivation of our prescription would be nice to have.
% 2) Can we compute anything concretely, even numerically in a three-matrix model?
% 3) Are there any useful ideas that we can think of in the context of matrix string theory?
% In this note, we have taken baby steps towards understanding bulk emergence in the BFSS model. In particular, we have demonstrated the features of target space entanglement that we believe must be present in any theory with a BFSS-like action. We did this by studying bosonic quantum mechanical models with two and three matrices. Studying entanglement in these models involved coming up with a gauge-invariant notion of computing entanglement entropies which has a clear geometric meaning in the bulk- It is the spatial entanglement between branes and strings separated by an entangling surface. 

% %Our results seem to suggest that the most natural notion of entanglement that can be defined in matrix quantum theory does not seem to have an Ryu-Takayanagi-like interpretation. Of course, we cannot rule out the possibility that this spatial entanglement computed in the bulk can be interpreted as the area of any cod-2 surface (in units of $1/( \hbar G_N)$.

% Drawing inspiration from EPR like entanglement we have demonstrated a connection between entanglement in lattice gauge theory and entanglement between matrix degrees of freedom. Even so, this is just a first step as there are many interesting questions which remain unanswered:

In this work we have discussed the computation of target space entanglement for gauged multimatrix models, in particular at weak coupling when the eigenvalues are well-separated.  There are many important directions for further research, of which we list a few below:

\paragraph{$N$ scaling}
A more complete and precise accounting of the $N$-scaling of target space entanglement entropy for two separated ``clumps" of eigenvalues is needed, in particular for understanding the holographic duals of this entropy in the situations studied for example in \cite{MollabashiNoburoTadashi:2014,KarchUhlemann,Andy:2017}.  

% We have demonstrated the $N$ scaling of the classical contribution to the entanglement entropy in the simple case of a maximally mixed state based upon string configurations between D0 branes. This is necessarily a state-dependent quantity. It would be interesting to understand in more generality how both the classical and quantum contributions scale as the number of degrees of freedom increases for a variety of states.

\paragraph{Entanglement and bulk surfaces} 
An obvious question is whether, in the limit of strong coupling and large $N$, the target space entanglement of matrix degrees of freedom has a dual holographic interpretation as the area of a bulk surface. This issue has been explored in several papers including \cite{MollabashiNoburoTadashi:2014,KarchUhlemann,Andy:2017,Anous_2019,Das:2020jhy,Das:2020xoa}, but there is as yet no conclusive answer. This is an area in which working with theories containing extended spatial directions, far out on the Coulomb branch, may be a useful strategy.

% The study of the full BFSS matrix model may lead to a better understanding of bulk emergence and a connection between entanglement between matrix degrees of freedom and bulk surfaces. This theory is of particular  interest because the dual field theory has no spatial degrees of freedom. As a result, studying the structure of entanglement between matrix degrees of freedom in this theory promises to teach us something fundamental about bulk emergence. Entanglement between the matrix degrees of freedom in the boundary theory might be associated to the areas of extremal surfaces that partition the compact dimensions in the bulk. However, a clear prescription relating a specific boundary quantity to the areas of these extremal surfaces is lacking. Of particular interest would be a connection between the work presented here and that of \cite{MollabashiNoburoTadashi:2014,KarchUhlemann} which both make use of the coulomb branch of a collection of D0 branes. It is also possible that extending our work to extended Dp branes, which would be interesting in its own right, may help facilitate this. We hope that the difficulties associated with defining a gauge-invariant notion of matrix entanglement in the gauge theory which we have addressed may give momentum towards a better understanding between entanglement in matrix models and bulk surfaces.

\paragraph{SUSY}
A more precise treatment of the fermionic degrees of freedom in a genuinely supersymmetric matrix model is an important direction for future work.

% 1) Straightforward extensions (?) D0-branes on a circle- the periodicity must show up in the entanglement entropy. What happens in the T-dual picture?
\paragraph{Dynamics}
%How do we set up a calculation to connect sub-AdS locality and entanglement between matrix degrees of freedom? 
In this paper, we have mainly focused on static entanglement measures in states with well separated D0-branes. It would be interesting to study entanglement dynamics in a scenario where D0-branes scatter off each other \cite{DKPS}. In the scattering region, the Born-Oppenheimer approximation breaks down and open strings stretching between D0-branes are created, introducing additional entanglement between groups of D0-branes.
%This is likely to help us understand how spacetime can be probed using entanglement. 
 
\paragraph{Multipartite entanglement} In this paper we have focused only on bipartite entanglement; it would be valuable to consider multipartite versions of target space entanglement in these models. 

\vskip .5cm
\centerline{\bf {\large Acknowledgements}}
\vskip .3cm

We would like to thank Aaron Fogel, Matthew Headrick, Djordje Radicevic, and Andrew Rolph for useful discussions on this paper and on related topics. HRH gratefully acknowledges the hospitality and support of Shirish and Shaila Lalwani during a period of quarantine when this work was completed. He also thanks Siddharth Mohite and Poortata Lalwani for helpful comments about the manuscript. This project was supported in part by a DOE HEP QuantISED grant DE-SC0020194, in part by DOE grant DE-SC0009987, and in part by the Simons Foundation through \emph{It from Qubit: Simons Collaboration on Quantum Fields, Gravity, and Information}.

%%%%%%%%%%%%%%%%%%%%%%%%%%

\appendix
\section{Derivation of the wavefunction and the measure} \label{MatrixMeasure}
The singlet condition implies that the wavefunction describing the D0-brane is invariant under an $U(N)$ transformation:
\be
\Psi'(X_i) = \Psi '(UXU^{\dagger}).
\ee 

We use this to diagonalize the matrix $X_1$. Let $X_1=U\La U^{\dagger}$ where we will denote the non-zero entries of the diagonal matrix $\La$ by  $\la_i, i=1 \cdots N$. The Jacobian of the transformation $(X_1, X_j)  \rightarrow (\La, U^{\dagger} X_j U), \quad j=2 \cdots N,$ is the square of the Vandermonde determinant,
\be
\Delta (\la_i) = \prod_{1 \leq i \leq j \leq N} (\la_i -\la_j).
\ee 

Moreover, the integral over $X_1$ involves an integration over the group manifold $SU(N)$ which we absorb in the wavefunction as follows,
\be
\Psi (\la_1 \cdots \la_N; X_2, \cdots X_N)= \sqrt{Vol[U(N)]} \Delta(\la_i)  \Psi'(\la_1 \cdots \la_N; X_2,X_3 \cdots X_N).
\ee 
 
We have successfully used the $U(N)$ gauge symmetry to diagonalize one of the matrices. This leaves a residual $U(1)^N \times S_N$ symmetry, where the permutation symmetry $S_N$ acts on the labels of each matrix element under which the wavefunction $\Psi'$ is invariant. Imposing this implies that $\Psi$ picks up the sign of the permutation. For instance, for a permutation $\s$,
\be
\Psi (\la_1 \cdots \la_N; X_2, \cdots X_N) = sign(\s) \Psi (\la_{\s(1)} \cdots \la_{\s(N)}; X_{\s(2)}, \cdots X_{\s(N)})
\ee 

Let us now specialize to the case of $N=2$ and two matrices. Expanding the state using Born-Oppenheimer approximation,
\be
|\psi \rangle = \int d x_1 d x_2 dy_1 dy_2 \sum_{n=0}^{\infty} [\psi_n(x_1,x_2,y_1,y_2) |x_1,x_2,y_1,y_2 \rangle |n;x_1-x_2 \rangle - (1 \leftrightarrow 2)]
\ee  
where we have denoted the diagonal entries of the two matrices $X$ and $Y$ with a single index and in small cases letters to emphasize the fact that we think of them as coordinates. The above expression can be easily generalized to the case of $N$ branes,
\be
|\psi \rangle = \int \prod_{i=1}^{i=N} dx_i dy_i|x_{i},y_{i}\rangle \Bigg( \prod_ {i,j=1 \cdots N}^{i<j} \sum_{n_{ij}=0}^{\infty} \psi_{n_{ij}}(x_{1} \cdots x_N;y_{1} \cdots y_N) | n_{ij}; x_i -x_j \rangle  \Bigg).
\ee 

It is understood that we need to take a tensor product of the both the fast and slow basis under the usual product. The harmonic oscillator states are normalized as usual $\langle n_{ij}; x_i-x_j | m_{kl} ; x_k -x_l \rangle = \delta_{ik} \delta_{jl} \delta_{m_{ij},n_{kl}}$ and thus, we can define a normalized state. 
	
\section{Corrections from nonadiabatic and finite coupling effects}\label{sec:BOCorrections}

In this Appendix, we estimate the size of non-adiabatic and finite-coupling corrections to the leading-order Born-Oppenheimer treatment, for low-energy states of the two-matrix model. We
will see that the natural small parameters are:
\begin{itemize}
    \item $\eps \equiv E_{slow}/E_{fast}$, measuring the strength
    of the adiabatic approximation. Here $E_{slow}$ is the energy of low-energy scattering states of the matrix eigenvalues, while $E_{fast}$ is the mass of excitations of the off-diagonal components.
    \item $g \equiv e^2 (\ell_s/\delta x)^3$, the strength of quartic interactions, where $\delta x$ is the separation between eigenvalues on each side of the entangling surface.
    \item We will also find corrections of order $\sqrt{g\eps}$.
\end{itemize} 
We take these parameters to be small, and show below that the corrections to the leading Born-Oppenheimer approximation are self-consistently of order $g,\eps,\sqrt{g\eps}$.

\subsection{Sub-leading Born-Oppenheimer corrections} 

%This is done by solving the Schrodinger equation for the fast modes  which leads to an effective Schrodinger equation for slow modes that can be solved. 

We write the $n =0$ and $n > 0$ terms in \eref{BO-wavefn} as: 
$\ket{\psi} = \ket{\psi_0} + \ket{\delta\psi_{n > 0}}$. The first term is the state constructed at lowest order in the Born-Oppenheimer approximation; the second term captures the ``non-adiabatic" corrections. We will consider solutions to the time-independent Schr\"odinger equation, assuming that the low-lying states lie in a continuum of scattering states. We will thus fix the low-lying energy; the goal of computing corrections to the Born-Oppenheimer approximation is to find a more accurate projection onto states with energies smaller than the masses of the ``fast" modes.

The time-independent Schr\"odinger equation  \eref{BO-Hamiltonian} acting on the first term in \eref{BO-wavefn} leads to:
\begin{eqnarray}
(H - E)\ket{\psi_0}
& = &  \frac{1}{\sqrt{2}} \int dx_1 dx_2 dy_1 dy_2 \Bigg[\Bigg( \Bigg( \frac{e^2 l_s}{2} (\pi^2_{x_1} +  \pi^2_{x_2} + \pi^2_{y_1} +\pi^2_{y_2})  + \frac{1}{l_s^2} |x_1-x_2| \Bigg) \nonumber\\ & &  \times \psi_0(x_1,x_2,y_1,y_2) | x_1,x_2,y_1,y_2 \rangle_s |0; x_1- x_2 \rangle_f \Bigg)-(1 \leftrightarrow 2)\Bigg] \nonumber \\ 
 & & \qquad - E |\psi_0 \rangle = - (H - E)\ket{\delta\psi}
\end{eqnarray}
Here, we have already taken $H_{fast}$ to lie in the instantaneous ground state; the term proportional to $|x_1 - x_2| \equiv \delta x$ in brackets is the eigenstate of $H_{fast}$ in Eq. (\ref{BO-Hamiltonian}). As discussed in the text, this term would spoil the assumption that we can have low-energy states for well-separated eigenvalues of $X$. In supersymmetric models such as \cite{BFSS:1997}, this vacuum energy is cancelled by fermions; while we do not include the fermions here, we will drop this term in the rest of our discussion. (The presence of fermions should not change our estimate of the size of corrections to the Born-Oppenheimer approximation, so this is the extent to which we will consider the effects of fermions here). Next, inserting resolutions of identity in the above equation, we find:
\begin{eqnarray}
& & \frac{-1}{\sqrt{2}} \int \prod_{i=1}
^{2} dx_i dy_i dz_i dw_i \Bigg( \psi_0(x_1,x_2,y_1,y_2)  |z_1,z_2,w_1,w_2 \rangle_s \nonumber\\
& & \qquad \qquad \times \langle w_2,w_1,z_2,z_1| \frac{e^2l_s}{2} (\del^2_{x_1} + \del^2_{x_2} + \del^2_{y_1} +\del^2_{y_2}) |x_1,x_2,y_1,y_2 \rangle_s |0;\delta x \rangle_f   - (1 \leftrightarrow 2) \Bigg) \nonumber\\
& & \qquad \qquad \qquad \qquad - E \ket{\psi_0} = - (H - E) |\delta \psi \rangle.
\end{eqnarray}
We can now use the fact that $ {}_s\langle z_i| -\del_{x}^2 | x_i \rangle_s = -\del_{x_i}^2 \delta^2 (x_i-z_i)$ and integrate by parts to find:
\begin{eqnarray} \label{leading_order BO}
& & \int \prod_{i=1}^{2}  \frac{dx_i dy_i}{\sqrt{2}}  \ket{x_i,y_i} \left[ \Big (-\frac{ e^2l_s}{2}(\del_{x_i}^2 +\del_{y_i}^2)\psi_0(x_i,y_i) - E\psi_0 \Big)  |0, \delta x \rangle 
\right.\nonumber\\
& & \qquad \qquad - e^2 l_s \left( \del_{x_i} \psi_0 \del_{x_i} + \del_{y_i} \psi_0 \del_{y_i} \right)\ket{0,\delta x}  \nonumber\\
& & \qquad \qquad \left. - \frac{e^2 l_s}{2}\psi_0(x_i,y_i) (\del_{x_i}^2 + \del_{y_i}^2) \ket{0,\delta x}-(1 \leftrightarrow 2) \right. \Big] = - (H - E)\ket{\delta \psi} , 
\end{eqnarray}
where $\del_{x_i}^2:= \del_{x_1}^2 +\del_{x_2}^2$, $\psi_0(x_i,y_i) :=\psi_0(x_1,x_2,y_1,y_2)$ and $\del_{x_i} \psi_0 \del_{x_i} :=\del_{x_1} \psi_0 \del_{x_1} + \del_{x_2} \psi_0 \del_{x_2} $. If we take the inner product of this equation with $\ket{x_i,y_i}_{antisym} \ket{0, \delta x} := \\ \frac{1}{\sqrt{2}}(\ket{x_1,x_2,y_1,y_2}\ket{0,\delta x}-(1 \leftrightarrow 2))$, the middle line gives the Berry connection. When the ground state wavefunction for the heavy modes is real, it is well known that this term then vanishes.\footnote{For a clear treatment of the Born-Oppenheimer approximation, including this point, see for example the papers in chapters 2 and 3 of \cite{wilczek1989geometric} and the many references cited therein.}
 The final term is of higher order in the Born-Oppenheimer approximation. To leading order, then, the wavefunction $\psi_0(x,y)$ satisfies the free wave equation. We can write
 these low-energy modes in terms of eigenstates of the target space
 momentum operator $-i \del_{x_i},-i\del_{y_i}$. If the characteristic wavenumber is $k$, the energy is $E_{slow} = e^2 \ell_s k^2/2$, characteristic of a free particle with mass $M = 1/(e^2 \ell_s)$.
 The Born-Oppenheimer approximation should be valid when $E_{slow} \ll E_{fast} \sim \delta x/l_s^2$
 
 Imposing this, the remaining terms are higher-order in the Born-Oppenheimer approximation (reflecting corrections to adiabaticity).
%where we have not included the term with derivatives acting on the fast basis as it is a higher order term. As discussed before, we assume that the zero-point energy due to the linear potential has been canceled by supersymmetry. We now look at the sub-leading and thus, the first non-trivial order to find,
Upon imposing the leading-order equations we are left with the subleading terms:
\begin{eqnarray} \label{sub-leading_order_BO}
& & \int \Big(\prod_{i=1}^{2} \frac{dx_i dy_i}{\sqrt{2}} \Big) \left[\ket{x_i,y_i} \Big( \frac{e^2l_s}{2}  \psi_0 (x_i,y_i)  (-\del_{x_i}^2 -\del_{y_i}^2) + e^2 l_s
 \left( \del_{x_i} \psi_0 \del_{x_i} + \del_{y_i} \psi_0 \del_{y_i} \right)\Big)\ket{0,\delta x}\right  .
\nonumber\\
& & \qquad + \sum_{n>0} \left. \Big(-\frac{e^2l_s}{2} (\del_{x_i}^2+\del_{y_i}^2) + n \frac{\delta x}{l_s^2} - E \Big) \delta \psi_n (x_i,y_i) |x_i,y_i \rangle |n;\delta x \rangle \right.- (1 \leftrightarrow 2)\Big] = 0.
\end{eqnarray}
 We wish to find $\delta \psi_n$ at the first subleading order in the corrections to adiabaticity. Thus, we ignore the terms in the second line of (\ref{sub-leading_order_BO}) for which the derivatives act on $\ket{n; \delta x}$. Taking the inner product of this equation with $\bra{\delta x, n}\bra{x_i,y_i}_{antisym}$, and solving formally for $\delta x$, we find:
\begin{eqnarray}
    & & \delta\psi_n = \frac{1}{\frac{e^2l_s}{2} (\del_{x_i}^2+\del_{y_i}^2) + n \frac{\delta x}{l_s^2} - E}\left(
    \frac{e^2l_s}{2} \psi_0(x_i,y_i) \bra{n,\delta x} (- \del_{x_i}^2 - \del_{y_i}^2)\ket{0,\delta x} \right.\nonumber\\
    & & \qquad\qquad \left. + e^2 l_s \left(\del_{x_i} \psi_0 
    \bra{n,\delta x} \del_{x_i} \ket{0,\delta x} + \del_{y_i} \psi_0
        \bra{n,\delta x} \del_{y_i} \ket{0,\delta x}\right)\right).
\end{eqnarray}
We can further simplify our estimate by replacing the denominator by its dominant term. We claim that to leading order the term $n \delta x/l_s^2 \sim E_{fast}$ dominates. It is larger than $E \equiv E_{slow}$ by a factor of $\eps$. The derivatives will act either on $\psi$ to bring down a factor of $k \sim \sqrt{2E_{slow}/e^2l_s}$ or on the matrix elements of the fast modes, bringing down a factor of $1/\delta x$. This gives terms of order $g,\eps,\sqrt{g \eps}$ relative to the dominant term. Thus, to leading order in these parameters, we find:
\begin{eqnarray}
    & & \delta\psi_n = \frac{1}{E_{fast}}\left(
    \frac{e^2l_s}{2} \psi_0(x_i,y_i) \bra{n,\delta x} (- \del_{x_i}^2 - \del_{y_i}^2)\ket{0,\delta x} \right.\nonumber\\
    & & \qquad\qquad \left. + e^2 l_s \left(\del_{x_i} \psi 
    \bra{n,\delta x} \del_{x_i} \ket{0,\delta x} + \del_{y_i} \psi_0
        \bra{n,\delta y} \del_{y_i} \ket{0,\delta x}\right)\right)
\end{eqnarray}
Applying the same principles as above to estimating the size of derivatives, the first term will give a correction of order $g\psi_0/n$, and the second, a correction of order $\sqrt{g\eps}\psi_0/n$.

\subsection{Perturbative corrections to ``fast" modes}\label{sec:pertcorr}

In computing the ground state of the fast modes for 3 or more matrices, we have in addition to the harmonic oscillator terms the commutator term ${\rm tr}[Y,Z]^2$ in Eq. (\ref{ThreeMatrixHamiltonian}).
This contributes the following classes of terms:
\begin{itemize}
    \item Additional contributions to the masses of off-diagonal terms:
    \be \delta H_{mass} = \frac{1}{e^2l_s^5} \sum_{i < j}\left[ (y_i - y_j)^2 |Z_{ij}|^2 + (z_i - z_j)^2 |Y_{ij}|^2\right]
    \ee
    \item Terms cubic in off-diagonal matrix elements, such as
    \be \Delta H_{cubic} = \frac{1}{2 e^2 l_s^5} \sum_{i < j < k}\left( 2 Y_{ii} - Y_{jj} - Y_{kk}\right) Z_{ij} Y_{jk} Z_{ki}
    \ee
    \item Terms quartic in off-diagonal matrix elements.
\end{itemize}
For simplicity, we will assume that all of the eigenvalues are well-separated by distances of order $\delta x \gg l_s$. Ignoring the cubic and quartic interactions, the off-diagonal modes have the Hamiltonian of simple harmonic oscillators with mass $M = 1/e^2l_s$ and frequencies $\omega \sim \delta x/l_s^2$. The characteristic spread of the wavefunction for the off-diagonal modes is
\be
    \delta Y = \sqrt{\vev{|Y_{ij}|^2}} \sim \delta Z 
    = \sqrt{\vev{|Z_{ij}|^2}}\sim \frac{1}{\sqrt{M \omega}} \sim \frac{e l_s^{3/2}}{\delta x^{1/2}} \equiv L
\ee
Recall that for a give perturbation $\delta H$ to the Hamiltonian, the ground state acquires a component proportional to the unperturbed excited state $\ket{n}$ with coefficient
\be
    \delta\ket{ground} \sim \frac{\bra{n}\delta H\ket{0}}{E_n - E_0} \ket{n}
\ee
where $\ket{0}$ is the ground states, and $E_n - E_0$ is the energy difference between the unperturbed energy eigenstates: here the energy difference is $\sim n\omega$. We estimate matrix elements of the form $\bra{n}ZYZ\ket{0} \sim L^2$, $\bra{n}ZYZ\ket{0} \sim L^4$. With this, the cubic interaction give corrections of order ${\cal O}(\sqrt{g})$, and the quartic interactions give corrections of order ${\cal O}(g)$

	\bibliographystyle{JHEP}
	\bibliography{mme}
	
\end{document}